\documentclass{article}

\usepackage{times}
\usepackage{graphicx} 
\usepackage{subfigure}

\usepackage{natbib}

\usepackage{algorithm}
\usepackage{algorithmic}
\usepackage[algo2e,linesnumbered,lined,boxed,vlined]{algorithm2e}

\usepackage{amsfonts}
\usepackage{amssymb}

\usepackage{hyperref}

\usepackage[accepted]{icml2016}

\usepackage{comment}
\usepackage{amsthm}
\usepackage{mathtools}
\usepackage{tabularx}
\usepackage{multirow}

\newtheorem{theorem}{Theorem}
\newtheorem{lemma}{Lemma}

\newtheorem{corollary}{Corollary}
\newtheorem{definition}{Definition}

\newtheorem{remark}{Remark}

\def\b{\ensuremath\boldsymbol}

\newcommand{\ind}{\perp\!\!\!\perp} 

\usepackage{lastpage}
\usepackage{fancyhdr}
\usepackage{url}
\icmltitlerunning{}

\begin{document}

\AddToShipoutPictureBG*{%
  \AtPageUpperLeft{%
    \setlength\unitlength{1in}%
    \hspace*{\dimexpr0.5\paperwidth\relax}
    \makebox(0,-0.75)[c]{\normalsize {\color{black} To appear as a part of an upcoming textbook on dimensionality reduction and manifold learning.}}
    }}

\twocolumn[
\icmltitle{Sufficient Dimension Reduction for High-Dimensional Regression and Low-Dimensional Embedding: Tutorial and Survey}

\icmlauthor{Benyamin Ghojogh}{bghojogh@uwaterloo.ca}
\icmladdress{Department of Electrical and Computer Engineering, 
\\Machine Learning Laboratory, University of Waterloo, Waterloo, ON, Canada}
\icmlauthor{Ali Ghodsi}{ali.ghodsi@uwaterloo.ca}
\icmladdress{Department of Statistics and Actuarial Science \& David R. Cheriton School of Computer Science, 
\\Data Analytics Laboratory, University of Waterloo, Waterloo, ON, Canada}
\icmlauthor{Fakhri Karray}{karray@uwaterloo.ca}
\icmladdress{Department of Electrical and Computer Engineering, 
\\Centre for Pattern Analysis and Machine Intelligence, University of Waterloo, Waterloo, ON, Canada}
\icmlauthor{Mark Crowley}{mcrowley@uwaterloo.ca}
\icmladdress{Department of Electrical and Computer Engineering, 
\\Machine Learning Laboratory, University of Waterloo, Waterloo, ON, Canada}

\icmlkeywords{Tutorial}

\vskip 0.3in
]

\begin{abstract}
This is a tutorial and survey paper on various methods for Sufficient Dimension Reduction (SDR). We cover these methods with both statistical high-dimensional regression perspective and machine learning approach for dimensionality reduction. We start with introducing inverse regression methods including Sliced Inverse Regression (SIR), Sliced Average Variance Estimation (SAVE), contour regression, directional regression, Principal Fitted Components (PFC), Likelihood Acquired Direction (LAD), and graphical regression. Then, we introduce forward regression methods including Principal Hessian Directions (pHd), Minimum Average Variance Estimation (MAVE), Conditional Variance Estimation (CVE), and deep SDR methods. Finally, we explain Kernel Dimension Reduction (KDR) both for supervised and unsupervised learning. We also show that supervised KDR and supervised PCA are equivalent. 
\end{abstract}

\section{Introduction}

Assume $X$ is the random variable of data and $Y$ is the random variables of labels of data. The labels can be discrete finite for classification or continuous for regression. 
Sufficient Dimensionality Reduction (SDR), first proposed in \cite{li1991sliced}, is a family of methods which find a transformation of data to a lower dimensional space which does not change the conditional of labels given data \cite{adragni2009sufficient}; hence, the subspace is sufficient for predicting labels from projected data onto the subspace.
This sufficient subspace is called the central subspace. 
The SDR can be divided into three main categories which are inverse regression methods, forward regression methods, and Kernel Dimension Reduction (KDR). In this paper, we introduce and explain various methods in these categories. The inverse and forward regression methods have a statistical approach while KDR has a machine learning approach mostly. 
Some surveys exist on SDR \cite{cook2001theory,chiaromonte2002sufficient,adragni2009sufficient,yin2011sufficient,ma2013review,li2018sufficient}.

It is noteworthy that the terminology ``dimension reduction" is often used in the literature of statistical regression while the terminology ``dimensionality reduction" is mostly used in the machine learning literature. Hence, the methods sufficient dimension reduction and kernel dimension reduction are sometimes called the sufficient dimensionality reduction and kernel dimensionality reduction, respectively. 

The remainder of this paper is organized as follows. 
We introduce the notations and preliminaries in Section \ref{section_preliminaries}.
Section \ref{section_inverse_regression_methods} introduces the inverse regression methods including Sliced Inverse Regression (SIR), Sliced Average Variance Estimation (SAVE), contour regression, directional regression, Principal Fitted Components (PFC), Likelihood Acquired Direction (LAD), and graphical regression. In Section \ref{section_forward_regression_methods}, we explain the forward regression methods such as Principal Hessian Directions (pHd), Minimum Average Variance Estimation (MAVE), Conditional Variance Estimation (CVE), and deep SDR methods. In Section \ref{section_KDR_methods}, we explain supervised and unsupervised KDR methods. Finally, Section \ref{section_conclusion} concludes the paper. 
Note that Sections \ref{section_inverse_regression_methods} and \ref{section_forward_regression_methods} have a statistical approach on regression while Sections \ref{section_deep_SDR} and \ref{section_KDR_methods} have a machine learning approach for dimensionality reduction. The reader can choose to skip some of the sections depending on their background but we recommend for readers with any background not to skip Section \ref{section_preliminaries}.

\section*{Required Background for the Reader}

This paper assumes that the reader has general knowledge of calculus, probability, and linear algebra. 

\section{Preliminaries and Notations}\label{section_preliminaries}


\begin{definition}[Regression problem]
Consider a dataset $\{(\b{x}_i, y_i)\}_{i=1}^n$ where $\{\b{x}_i\}_{i=1}^n$ are called the covariates, features, explanatory, or predictor variables and $\{y_i\}_{i=1}^n$ are called the labels, targets, or responses. 
The dimensionality of covariates and labels are $\b{x}_i \in \mathbb{R}^d$ and $y_i \in \mathbb{R}$. 
The labels can be either continuous or discrete finite. In the latter case, the problem is a classification problem where every label states which class the point belongs to. 
A regression problem relates every covariate $\b{x}$ and its corresponding label $y$ as \cite{li1991sliced}:
\begin{align}\label{equation_regression_problem_vector}
y = f(\b{U}^\top \b{x}, \b{\varepsilon}) = f(\b{u}_1^\top \b{x}, \b{u}_2^\top \b{x}, \dots, \b{u}_p^\top \b{x}, \b{\varepsilon}),
\end{align}
where $\b{U} = [\b{u}_1, \dots, \b{u}_p] \in \mathbb{R}^{d \times p}$ is the projection matrix which projects data onto its $p$-dimensional column space (with $p \leq d$), $f: \mathbb{R}^{p+1} \rightarrow \mathbb{R}$ is an arbitrary unknown function, and $\b{\varepsilon}$ is some scalar noise independent of $\b{x}$.
Writing Eq. (\ref{equation_regression_problem_vector}) in matrix form gives:
\begin{align}\label{equation_regression_problem_matrix}
\b{y} = f(\b{U}^\top \b{X}, \b{\varepsilon}),
\end{align}
where $\b{y} = [y_1, \dots, y_n]^\top \in \mathbb{R}^n$ and $\b{X} = [\b{x}_1, \dots, \b{x}_n] \in \mathbb{R}^{d \times n}$.
Note that some papers consider a regression problem to be:
\begin{align}\label{equation_regression_problem_vector_2}
y = f(\b{U}^\top \b{x}) + \varepsilon = f(\b{u}_1^\top \b{x}, \b{u}_2^\top \b{x}, \dots, \b{u}_p^\top \b{x}) + \varepsilon,
\end{align}
where $f: \mathbb{R}^{p} \rightarrow \mathbb{R}$ is an arbitrary unknown function.
Note that the projection $\b{U}^\top \b{x}$ is a linear projection \cite{cunningham2015linear}.
\end{definition}

\begin{remark}\label{remark_row_wise_regression}
Some papers on regression take the $\b{X}$ axis row-wise to have $\b{X} = [\b{x}_1, \dots, \b{x}_n]^\top \in \mathbb{R}^{n \times d}$. In this case, the regression problem is:
\begin{align}\label{equation_regression_problem_matrix_rowWise}
\b{y} = f(\b{X} \b{U}, \b{\varepsilon}), \quad \text{ or }\quad \b{y} = f(\b{X} \b{U}) + \varepsilon.
\end{align}
In this paper, we take the default for $\b{X}$ to be column-wise unless otherwise specified for some algorithms.
\end{remark}

We denote the random variables associated with the covariates $\b{X}$ and labels $\b{y}$ by $X$ and $Y$ denote, respectively. 

\begin{remark}[Relation to projection pursuit regression]
The model of Eq. (\ref{equation_regression_problem_vector_2}) is also used in Projection Pursuit Regression (PPR) \cite{friedman1981projection}. Also, one can refer to \cite{friedman1974projection} for reading about Projection Pursuit (PP).
\end{remark}

\begin{definition}[Effective dimension reduction (e.d.r.) space {\citep[Section 1]{li1991sliced}}]\label{definition_edr_subspace}
Eq. (\ref{equation_regression_problem_vector}) or (\ref{equation_regression_problem_matrix}) shows that the $d$-dimensional covariates are projected onto a $p$-dimensional subspace, polluted with some noise, and then fed to some unknown function to give us the labels. This subspace is called the Effective Dimension Reduction (e.d.r.) space.
As $\b{U}$ is the projection matrix, its columns are bases for the e.d.r. subspace. Hence, in some papers on regression, this matrix is denoted by $\b{B}$ to denote bases. The vectors $\{\b{u}_i\}_{i=1}^p$ (also denoted by $\{\b{\beta}_i\}_{i=1}^p$ in the literature) are the bases for the e.d.r. subspace.
\end{definition}

\begin{definition}[Dimension reduction subspace and the central subspace {\citep[Section 3]{cook2000save}}]\label{definition_central_subspace}
Consider a reduction function $P_\mathcal{S}: \mathbb{R}^d \rightarrow \mathbb{R}^p$ where $p \leq d$.
Let projection of $\b{X}$ onto a subspace $\mathcal{S}$ be denoted by $P_\mathcal{S} \b{X}$ (e.g., it is $P_\mathcal{S} \b{X} = \b{U}^\top \b{X}$ in Eq. (\ref{equation_regression_problem_vector})).
The subspace is called the dimension reduction subspace if we have \cite{li1991sliced,li1992principal}:
\begin{align}\label{equation_X_Y_independent_projection}
X \ind Y\, |\, P_\mathcal{S} X,
\end{align}
where $\ind$ denotes independence of random variables \cite{dawid1979conditional} and $|$ denotes conditioning. 
Considering $P_\mathcal{S} X = \b{U}^\top X$, Eq. (\ref{equation_X_Y_independent_projection}) can be restated as {\citep[Section 4]{cook1994interpretation}}:
\begin{align}\label{equation_X_Y_independent_projection_2}
X \ind Y\, |\, \b{U}^\top X.
\end{align}
We desire to find the smallest dimension reduction subspace with minimal dimension, called the minimal subspace. This smallest subspace may not be unique {\citep[pp. 104-105]{cook1998regression}}. 
The central dimension reduction subspace, or called the central subspace in short, is denoted by $\mathcal{D}_{Y|X}$ and is defined to be the intersection of all dimension reduction subspaces for regression \cite{cook1994using,cook1996graphics}. Some weak conditions are required for the central subspace to exist (cf. \cite{cook1996graphics,cook1998regression}).
\end{definition}

\begin{definition}[Central mean subspace \cite{cook2002dimension}]
The subspace spanned by $\mathbb{E}[\b{x} | y]$ is called the central mean subspace, denoted by $\mathcal{S}_{\mathbb{E}[\b{x} | y]}$.
The central mean subspace is always contained in the central subspace \cite{li2005contour}. 
\end{definition}

\begin{definition}[Sufficient reduction \cite{cook2007fisher}, {\citep[Definition 1.1]{adragni2009sufficient}}]\label{definition_sufficient_reduction}
A reduction or projection onto subspace $P_\mathcal{S}: \mathbb{R}^d \rightarrow \mathbb{R}^p$ (where $p \leq d$) is sufficient if it satisfies at least one of the followings:
\begin{itemize}
\item inverse reduction: $X\, |\, (Y, P_\mathcal{S} X)$ is identically distributed as $X\, |\, P_\mathcal{S} X$.
\item forward reduction: $Y\, |\, X$ is identically distributed as $Y\, |\, P_\mathcal{S} X$.
\item joint reduction: $X \ind Y\, |\, P_\mathcal{S} X$ which is Eq. (\ref{equation_X_Y_independent_projection}).
\end{itemize}
\end{definition}

\begin{definition}[Effective subspace for regression \cite{fukumizu2003kernel}]\label{definition_effective_subspace}
Let $P_{\mathcal{S}}\b{X} = \b{U}^\top \b{X}$ be projection onto a $p$-dimensional subspace. 
Let $\mathbb{R}^{d \times d} \ni \b{Q} = [\b{U} | \b{V}]$ be an orthogonal matrix, where $\b{U} \in \mathbb{R}^{d \times p}$ is the truncated projection matrix onto the $p$-dimensional subspace and $\b{V} \in \mathbb{R}^{d \times (d-p)}$ is the rest of matrix $\b{Q}$.
If the subspace $\mathcal{S}$ is an effective subspace for regression, we have the following relations for the conditional probabilities:
\begin{align}
&\mathbb{P}_{Y | P_{\mathcal{S}}X}(\b{y}\, |\, P_{\mathcal{S}}\b{X}) = \mathbb{P}_{Y | X}(\b{y}\, |\, \b{X}), \label{equation_effective_subspace_1} \\
&\mathbb{P}_{Y | \b{U}^\top X, \b{V}^\top X}(\b{y}\, |\, \b{U}^\top \b{X}, \b{V}^\top \b{X}) = \mathbb{P}_{Y | \b{U}^\top X}(\b{y}\, |\, \b{U}^\top \b{X}), \label{equation_effective_subspace_2}
\end{align}
which are equivalent equations.
Both equations mean that all the required information for predicting the labels $\b{Y}$ are contained in the subspace $\mathcal{S}$ which is the column space of $\b{U}$; hence, this subspace is sufficient for $\b{X}$ to be projected onto. 
We also have:
\begin{align*}
I(Y,X) = &\,I(Y,\b{U}^\top X) \\
&+ \mathbb{E}_{\b{U}^\top X}[I(Y | \b{U}^\top X, \b{V}^\top X | \b{U}^\top X)],
\end{align*}
where $I(\cdot|\cdot)$ is the mutual information between random variables.
\end{definition}

The e.d.r. subspace (Definition \ref{definition_edr_subspace}), the central subspace (Definition \ref{definition_central_subspace}), and effective subspace for regression (Definition \ref{definition_effective_subspace}) have equivalent meanings. 
The e.d.r. subspace, central subspace, and effective subspace are often used in Prof. Ker-Chau Li's papers, Prof. R. Dennis Cook's papers, and the literature of kernel dimension reduction, respectively. 

\begin{definition}[Exhaustive dimension reduction {\citep[Section 3]{li2007directional}}]\label{definition_exhaustive_method}
A dimension reduction method estimates a subspace $\mathcal{S}$ of $\mathcal{S}_{Y|X}$, i.e., $\mathcal{S} \subseteq \mathcal{S}_{Y|X}$. If we have $\mathcal{S} = \mathcal{S}_{Y|X}$, the method is called to be exhaustive.
\end{definition}

\begin{definition}[Linear regression or OLS problem]\label{definition_linear_regression}
A linear regression problem, also called the Ordinary Least Squares (OLS) problem, is a special case of Eq. (\ref{equation_regression_problem_matrix_rowWise}) which usually considers $\b{X}$ row-wise. It takes the function $f$ to be an identity function so we have:
\begin{align*}
\b{Y} = \b{X} \b{U} + \b{\varepsilon},
\end{align*}
where $\b{X} \in \mathbb{R}^{n \times d}$, $\b{U} \in \mathbb{R}^{d \times p}$ and the labels can be multi-dimensional, i.e., $\b{Y} \in \mathbb{R}^{n \times p}$. 
The projection matrix, or so-called coefficients, $\b{U}$ can be calculated using a least squares problem:
\begin{align}
&\underset{\b{U}}{\text{min.}}\,\, \|\b{Y} - \b{X} \b{U}\|_F^2 \nonumber\\
&\implies \frac{\partial }{\partial \b{U}}  \|\b{Y} - \b{X} \b{U}\|_F^2 = -2\b{X}^\top(\b{Y} - \b{X} \b{U}) \overset{\text{set}}{=} \b{0} \nonumber\\
&\implies \b{X}^\top \b{Y} = \b{X}^\top \b{X} \b{U} \implies \b{U} = (\b{X}^\top \b{X})^{-1} \b{X}^\top \b{Y}, \label{equation_linear_regression_U_matrix}
\end{align}
where $\|.\|_F$ denotes the Frobenius norm.
There also exist some other weighted methods for linear regression such as Iteratively Re-weighted Least Squares (IRLS) \cite{chartrand2008iteratively}.
\end{definition}

Let the mean and covariance of covariates be denoted by $\mathbb{E}[\b{x}]$ (or $\b{\mu}_x$) and $\b{\Sigma}_{xx}$, respectively.
We can standardize the covariates to have:
\begin{align}\label{equation_z}
\b{z} := \b{\Sigma}_{xx}^{-1/2} (\b{x} - \mathbb{E}[\b{x}]).
\end{align}
We denote the random variable associated with $\b{z}$ by $Z$.
We can restate Eq. (\ref{equation_regression_problem_vector}) as:
\begin{align}\label{equation_regression_problem_vector_standardized}
y = f(\b{W}^\top \b{z}, \b{\varepsilon}) = f(\b{w}_1^\top \b{z}, \b{w}_2^\top \b{z}, \dots, \b{w}_p^\top \b{z}, \b{\varepsilon}),
\end{align}
where $\b{W} = [\b{w}_1, \dots, \b{w}_p] = \b{\Sigma}_{xx}^{1/2} \b{U} = [\b{\Sigma}_{xx}^{1/2} \b{u}_1, \dots, \b{\Sigma}_{xx}^{1/2} \b{u}_p] \in \mathbb{R}^{d \times p}$ is the projection matrix for the standardized data. 
The $\b{w}_1, \dots, \b{w}_p$ are called the standardized projection directions. 

\begin{corollary}[Relation of subspace bases for covariates and standardized covariates {\citep[Section 2]{li1991sliced}}]
According to Eqs. (\ref{equation_regression_problem_matrix}), (\ref{equation_z}), and (\ref{equation_regression_problem_vector_standardized}), the relation between the projection directions (i.e. bases for covariates) and the standardized projection directions (i.e., bases for standardized covariates) are:
\begin{align}\label{equation_u_w_relation}
\b{w}_i = \b{\Sigma}_{xx}^{1/2} \b{u}_i \implies \b{u}_i = \b{\Sigma}_{xx}^{-1/2} \b{w}_i, \quad \forall i \in \{1, \dots, p\}.
\end{align}
\end{corollary}

\begin{corollary}[Relation of central subspaces for covariates and standardized covariates {\citep[Section 4]{cook2000save}}]
According to Eq. (\ref{equation_u_w_relation}), the relation of central subspaces for covariates and standardized covariates is:
\begin{align}
\mathcal{S}_{Y|Z} = \b{\Sigma}_{xx}^{1/2} \mathcal{S}_{Y|X}.
\end{align}
\end{corollary}

We can estimate the mean and covariance of covariates with the sample mean and sample covariance matrix, respectively:
\begin{align}
& \mathbb{E}[\b{x}] \approx \widehat{\b{\mu}}_x := \frac{1}{n} \sum_{i=1}^n \b{x}_i, \label{equation_sample_mean} \\
& \b{\Sigma}_{xx} \approx \widehat{\b{\Sigma}}_{xx} := \frac{1}{n} \sum_{i=1}^n (\b{x}_i - \bar{\b{x}}) (\b{x}_i - \bar{\b{x}})^\top. \label{equation_sample_covariance}
\end{align}
We can use these estimates in Eq. (\ref{equation_z}) to estimate the standardized covariates:
\begin{align}\label{equation_z_i_hat}
\mathbb{R}^{d} \ni \widehat{\b{z}}_i := \widehat{\b{\Sigma}}_{xx}^{-1/2} (\b{x}_i - \widehat{\b{\mu}}_x).
\end{align}
Also, according to Eq. (\ref{equation_u_w_relation}), we have:
\begin{align}\label{equation_u_w_relation_estimate}
\b{u}_i = \widehat{\b{\Sigma}}_{xx}^{-1/2} \b{w}_i, \quad \forall i \in \{1, \dots, p\}.
\end{align}

Let $X'$, $Y'$, and $Z'$ be independent copies of random variables $X$, $Y$, and $Z$, respectively. 

\begin{definition}[Stiefel and Grassmannian manifolds \cite{absil2009optimization}]\label{definition_Stiefel_Grassmannian_manifolds}
The Stiefel manifold is defined as the set of orthogonal matrices as:
\begin{align}
\mathcal{S}t(p,d) := \{\b{U} \in \mathbb{R}^{d \times p}\, |\, \b{U}^\top \b{U} = \b{I}\},
\end{align}
where $p \leq d$.
The Grassmannian manifold $\mathcal{G}(p,d)$ is defined to be all $p$-dimensional subspaces of $\mathbb{R}^d$. The Grassmannian manifold can be seen as the quotient space of the Stiefel manifold $\mathcal{S}t(p,d)$ denoted as:
\begin{align}
\mathcal{G}(p,d) := \mathcal{S}t(p,d) / \mathcal{S}t(p,p).
\end{align}
The reader can refer to \cite{absil2009optimization} to know more about quotient spaces. 
The Stiefel and Grassmannian manifolds are examples for Riemannian manifold which is a smooth manifold endowed with a metric. 
\end{definition}
The projection matrix $\b{U}$ onto the central subspace is usually an orthogonal matrix because the bases of the subspace, which are the columns of the projection matrix, are orthonormal vectors. 
Therefore, according to Definition \ref{definition_Stiefel_Grassmannian_manifolds}, the projection matrix of a central subspace belongs to the Stiefel manifold. Moreover, as central space is a subspace of the space of covariates, it belongs to Grassmannian manifold, according to Definition \ref{definition_Stiefel_Grassmannian_manifolds}.
Many SDR methods such as LAD \cite{cook2009likelihood} and CVE \cite{fertl2021conditional} use Riemannian optimization \cite{absil2009optimization} to find the central subspace in these manifolds \cite{cunningham2015linear}. 

\section{Inverse Regression Methods}\label{section_inverse_regression_methods}

\begin{definition}[Inverse regression problem]\label{definition_inverse_regression}
A forward regression problem tries to estimate the labels from the covariates by calculating the bases of central subspace. A reverse regression problem exchanges the roles of covariates and labels and tries to estimate the covariates from the labels by calculating the bases of central subspace.
\end{definition}

Inverse regression methods are based on inverse regression. Some important inverse regression methods are Sliced Inverse Regression (SIR) \cite{li1991sliced}, Sliced Average Variance Estimation (SAVE) \cite{cook1991sliced,cook2000save}, Parametric Inverse Regression (PIR) \cite{bura2001estimating}, Contour Regression (CR) \cite{li2005contour}, Directional Regression (DR) \cite{li2007directional}, Principal Fitted Components (PFC) \cite{cook2007fisher,cook2008principal}, Likelihood Acquired Direction (LAD) \cite{cook2009likelihood}, and graphical regression \cite{cook1998regression,cook1998regressionPaper}. There are some other methods such as SDR by inverse of the $k$-th moment \cite{yin2003estimating}, SDR by minimum discrepancy \cite{cook2005sufficient}, SDR by Intra-slice covariances \cite{cook2006using}, SDR for non-elliptically distributed covariates \cite{li2009dimension}, envelope models \cite{cook2010envelope}, and Principal Support Vector Machines (PSVM) \cite{li2011principal}, which are not covered for brevity. 
The inverse regression methods usually have strict assumptions on the distributions and conditional distributions. 

\begin{definition}[The inverse regression curve {\citep[Section 3]{li1991sliced}}]
The $d$-dimensional inverse regression curve is defined as
$\mathbb{E}[\b{x} | y]$ versus $y$. The centered inverse regression curve is $\mathbb{E}[\b{x} | y] - \mathbb{E}[\b{x}]$ versus $y$. 
Hence, the standardized inverse regression curve is
$\mathbb{E}[\b{z} | y]$ versus $y$.
\end{definition}

\begin{lemma}[Linearity condition \cite{eaton1986characterization}]\label{lemma_linearity_condition}
The covariates have elliptical distribution if and only if for any $\b{b} \in \mathbb{R}^d$, the conditional expectation is linear, i.e., $\mathbb{E}[\b{b}^\top \b{x}\, |\, \b{u}_1^\top \b{x}, \dots, \b{u}_p^\top \b{x}] = c_0 + c_1 \b{u}_1^\top \b{x} + \dots + c_p \b{u}_p^\top \b{x}$. This is called the linearity condition. 
\end{lemma}

The linearity condition is assumed on almost all inverse regression methods. 

\begin{remark}[On linearity condition]
The linearity condition holds most often \cite{li1991sliced} because projection of high dimensional covariates onto a low dimensional subspace has a distribution close to normal distribution \cite{diaconis1984asymptotics,hall1993almost}.
\end{remark}

\subsection{Sliced Inverse Regression (SIR)}

Sliced Inverse Regression (SIR) was first proposed in \cite{li1991sliced}. We explain it in the following. 

\begin{theorem}[{\citep[Condition 3.1, Theorem 3.1, and Corollary 3.1]{li1991sliced}}]\label{theorem_SIR_inverseRegressionCurve_contained_in_subspace}
Assume the linearity condition in Lemma \ref{lemma_linearity_condition} holds.
The centered inverse regression curve $\mathbb{E}[\b{x} | y] - \mathbb{E}[\b{x}]$ is contained in the linear subspace spanned by $\{\b{\Sigma}_{xx} \b{u}_1, \dots, \b{\Sigma}_{xx} \b{u}_p\}$.
Moreover, in this case, the standardized inverse regression curve $\mathbb{E}[\b{z} | y]$ is contained in the linear subspace spanned by $\{\b{w}_1, \dots, \b{w}_p\}$.
\end{theorem}

\begin{definition}[Slicing labels in regression \cite{li1991sliced}]
Let the range of labels be denoted by set $\mathcal{D}_y$.
We can divide the range of labels into $h$ slices, i.e. $\{\mathcal{D}_1, \dots, \mathcal{D}_h\}$, where:
\begin{align*}
& \bigcup_{s=1}^h \mathcal{D}_s = \mathcal{D}_y, \\
& \mathcal{D}_{s_1} \cap \mathcal{D}_{s_2} = \varnothing, ~~~ \forall s_1,s_2 \in \{1, \dots, h\}, ~ s_1 \neq s_2.
\end{align*}
Let the proportion of labels which fall in the slice $\mathcal{D}_s$ be denoted by $\rho_s$. We have $\rho_1 + \dots + \rho_h = 1$. We use the proportions because the slices may not be equally populated. 
\end{definition}

If the regression problem is actually a classification problem whose labels are discrete finite, slicing is not required as the labels are already sliced into the discrete finite values. In that case, the number of slices $h$ is equal to the number of classes. 

\begin{lemma}[Central subspace for binary labels {\citep[Section 4]{cook2000save}}]\label{lemma_central_subspace_binary_labels}
Let the labels become binary as:
\begin{align*}
\widetilde{y} := 
\left\{
    \begin{array}{ll}
        1 & \mbox{if } y > c \\
        0 & \mbox{if } y \leq c,
    \end{array}
\right.
\end{align*}
for some constant $c$ and let its associated random variable be $\widetilde{Y}$. We have:
\begin{align}\label{equation_Y_binary_subspace_subset}
\mathcal{S}_{\widetilde{Y}|X} \subseteq \mathcal{S}_{Y|X}.
\end{align}
\end{lemma}

\begin{corollary}[Relation of central subspaces for labels and sliced labels {\citep[Section 4]{cook2000save}}]\label{corollary_sliced_Y_subspace_relation}
An extension of Lemma \ref{lemma_central_subspace_binary_labels} is as follows. Let $\widetilde{Y}$ denote the random variable for sliced labels into $h$ slices, where $\widetilde{y}_i = k$ if $y_i \in \mathcal{D}_k$. Then, Eq. (\ref{equation_Y_binary_subspace_subset}) holds for the sliced labels. 
\end{corollary}

We compute the sample mean of standardized covariates whose labels fall in every slice:
\begin{align}\label{equation_mean_of_slice}
\mathbb{R}^{d} \ni \widehat{\b{\mu}}_{z,s} := \frac{1}{n \rho_s} \sum_{y_i \in \mathcal{D}_s} \widehat{\b{z}}_i.
\end{align}
Then, we calculate the Principal Component Analysis (PCA) of the points $\{\widehat{\b{\mu}}_{z,s}\}_{s=1}^h$. The PCA of these points can be done by eigenvalue decomposition of their weighted covariance matrix \cite{ghojogh2019unsupervised}. The weighted covariance matrix is \cite{li1991sliced}:
\begin{align*}
\mathbb{R}^{d \times d} \ni \widehat{\b{V}} := \sum_{s=1}^h \rho_s\, \widehat{\b{\mu}}_{z,s}\, \widehat{\b{\mu}}_{z,s}^\top.
\end{align*}
We consider the top $p$ eigenvectors of $\widehat{\b{V}}$ with largest eigenvalues \cite{ghojogh2019eigenvalue}.
According to Theorem \ref{theorem_SIR_inverseRegressionCurve_contained_in_subspace}, these eigenvectors are $\{\b{w}_j\}_{j=1}^p$ used in Eq. (\ref{equation_regression_problem_vector_standardized}). 
Finally, we use Eq. (\ref{equation_u_w_relation_estimate}) to calculate $\{\b{u}_j\}_{j=1}^p$ from $\{\b{w}_j\}_{j=1}^p$. The calculated $\{\b{u}_j\}_{j=1}^p$ are the bases for the central subspace and are used in Eq. (\ref{equation_regression_problem_vector}).
According to Eq. (\ref{equation_Y_binary_subspace_subset}), SIR is not an exhaustive method (see Definition \ref{definition_exhaustive_method}). 
Note that SIR has also been extended to work for categorical (i.e., discrete finite) covariates \cite{chiaromonte2002sufficient2}.

\subsection{Sliced Average Variance Estimation (SAVE)}\label{section_SAVE}

Sliced Average Variance Estimation (SAVE) was first proposed in \cite{cook2000save} while it was developed over \cite{cook1991sliced,cook1999dimension}.
Two assumptions are required for SAVE \cite{cook2000save}:
\begin{enumerate}
\item Linearity condition: As in Lemma \ref{lemma_linearity_condition}, we assume that $\mathbb{E}[Z | \b{U}^\top Z]$ is linear in $\b{U}^\top Z$ resulting in $\mathbb{E}[Z | \b{U}^\top Z] = P_{\mathcal{S}_{Y|Z}} Z$ which is equivalent to $Z$ having an elliptical distribution \cite{eaton1986characterization}. 
\item Constant covariance condition: We also assume that $\mathbb{V}\text{ar}[Z | \b{U}^\top Z]$ is constant, which is equivalent to $Z$ having a normal distribution \cite{cook2000save}. 
We slice the labels into $h$ slices.
\end{enumerate}

\begin{theorem}[{\citep[Section 5]{cook2000save}}]\label{theorem_SAVE_span}
Let $\text{span}\{\cdots\}$ denote a space spanned by the set of bases. We have:
\begin{align*}
\text{span}\Big\{\mathbb{E}\big[\b{I} - \mathbb{V}\text{ar}[Z | \widetilde{Y}]\big]^2\Big\} \subseteq \mathcal{S}_{\widetilde{Y}|X} \overset{(\ref{equation_Y_binary_subspace_subset})}{\subseteq} \mathcal{S}_{Y|X},
\end{align*}
where $\widetilde{Y}$ is the random variable for sliced label, defined in Corollary \ref{corollary_sliced_Y_subspace_relation}.
\end{theorem}

The sample covariance for every slice $s$ is:
\begin{align*}
\mathbb{R}^{d \times d} \ni \widehat{\b{V}}_s := \frac{1}{n \rho_s} \sum_{y_i \in \mathcal{D}_s} (\widehat{\b{z}}_i - \widehat{\b{\mu}}_{z,s}) (\widehat{\b{z}}_i - \widehat{\b{\mu}}_{z,s})^\top,
\end{align*}
where $\widehat{\b{\mu}}_{z,s}$ is defined in Eq. (\ref{equation_mean_of_slice}). This $\widehat{\b{V}}_s$ is an estimation for $\mathbb{V}\text{ar}[Z | \widetilde{Y}]$.
Consider the following matrix:
\begin{align*}
\mathbb{R}^{d \times d} \ni \b{M} := \sum_{s=1}^h \rho_s (\b{I} - \widehat{\b{V}}_s)^2,
\end{align*}
where $\b{I}$ is the identity matrix. 
We consider the top $p$ eigenvectors of $\b{M}$ with largest eigenvalues \cite{ghojogh2019eigenvalue}.
According to Theorem \ref{theorem_SAVE_span}, these eigenvectors are $\{\b{w}_j\}_{j=1}^p$ used in Eq. (\ref{equation_regression_problem_vector_standardized}). 
We use Eq. (\ref{equation_u_w_relation_estimate}) to calculate $\{\b{u}_j\}_{j=1}^p$ from $\{\b{w}_j\}_{j=1}^p$. The calculated $\{\b{u}_j\}_{j=1}^p$ are the bases for the central subspace and are used in Eq. (\ref{equation_regression_problem_vector}).
SAVE is an exhaustive method (see Definition \ref{definition_exhaustive_method}), according to \cite{li2007directional}.

\subsection{Parametric Inverse Regression (PIR)}

PIR was first proposed in \cite{bura2001estimating}. PIR considers the row-wise regression as in Remark \ref{remark_row_wise_regression}. Eq. (\ref{equation_regression_problem_matrix_rowWise}) is a forward regression problem. We consider the inverse regression problem (see Definition \ref{definition_inverse_regression}) for the standardized covariates. PIR allows the labels to be multi-dimensional ($\ell$-dimensional), i.e., $\b{Y} \in \mathbb{R}^{n \times \ell}$.
\begin{align*}
\b{Z} = f(\b{Y})\, \b{B} + \b{\varepsilon},
\end{align*}
where $\b{Z} \in \mathbb{R}^{n \times d}$ are the standardized covariates, $f: \mathbb{R}^{n \times \ell} \rightarrow \mathbb{R}^{n \times \ell}$ is some function, $\b{B} \in \mathbb{R}^{\ell \times p}$ is the projection matrix onto a $p$-dimensional subspace, and $\epsilon$ denotes independent noise.
According to Eq. (\ref{equation_linear_regression_U_matrix}), we can calculate $\b{B}$ as:
\begin{align*}
\b{B} = \big(f(\b{Y})^\top f(\b{Y})\big)^{-1} f(\b{Y})^\top \b{Z},
\end{align*}
which is the projection matrix onto a $p$-dimensional central subspace. 

\subsection{Contour Regression (CR)}

Contour Regression (CR) was proposed in \cite{li2005contour} with two versions which are simple and general contour regression. Here, we introduce the simple contour regression. 
CR claims that all the directional information of covariates exists in the set $\{\b{x}_i - \b{x}_j\, |\, 1 \leq i < j \leq n\}$.

We assume the covariates have an elliptical distribution (see Lemma \ref{lemma_linearity_condition}). 
We also assume that {\citep[Assumption 2.1]{li2005contour}}:
\begin{align*}
&\mathbb{V}\text{ar}\big[\b{w}^\top (Z' - X)\, \big|\, |Y'-Y| \leq c\big] > \\
&~~~~~~~~~~~~~~~~~\mathbb{V}\text{ar}\big[\b{v}^\top (Z' - X)\, \big|\, |Y'-Y| \leq c\big],
\end{align*}
where $\b{v} \in \mathcal{S}_{Y|Z}$, $\b{w} \in (\mathcal{S}_{Y|Z})^{\bot}$, $\|\b{v}\|_2 = \|\b{w}\|_2 = 1$, and $c > 0$. Note that $\mathcal{S}^{\bot}$ denotes the orthogonal space to $\mathcal{S}$.

\begin{theorem}[{\citep[Theorem 2.1 and Corollary 2.1]{li2005contour}}]\label{theorem_CR}
Let the above assumptions hold. Consider the matrix:
\begin{equation}\label{equation_CR_M_matrix}
\begin{aligned}
\b{M} := \mathbb{E}\big[(X' - X)(X' - X)^\top\, \big|\, |Y' - Y| \leq c\big],
\end{aligned}
\end{equation}
with $c > 0$.
We consider the $p$ tailing eigenvectors of $\b{\Sigma}_{xx}^{-1/2} \b{M} \b{\Sigma}_{xx}^{-1/2}$ with smallest eigenvalues \cite{ghojogh2019eigenvalue}, denoted by $\gamma_{d-p+1}, \dots, \gamma_{d}$. The $\{\b{u}_j\}_{j=1}^p$ used in Eq. (\ref{equation_regression_problem_vector}) are the vectors $\b{\Sigma}_{xx}^{-1/2} \gamma_{d-p+1}, \dots, \b{\Sigma}_{xx}^{-1/2} \gamma_{d}$.
\end{theorem}

We can estimate Theorem \ref{theorem_CR} in practice as follows. 
The Eq. (\ref{equation_CR_M_matrix}) is estimated as \cite{li2005contour}:
\begin{equation*}
\begin{aligned}
\widehat{\b{M}} = \frac{1}{\binom{n}{c}} \sum_{i=1}^n \sum_{j=1}^n (\b{x}_j - \b{x}_i)(\b{x}_j - \b{x}_i)^\top \mathbb{I}(|\b{y}_j - \b{y}_i| \leq c),
\end{aligned}
\end{equation*}
where $\mathbb{I}(.)$ is the indicator function which is one if its condition is satisfied and is zero otherwise. 
We can compute the $p$ tailing eigenvectors of $\widehat{\b{\Sigma}}_{xx}^{-1/2} \widehat{\b{M}} \widehat{\b{\Sigma}}_{xx}^{-1/2}$ where the sample covariance is defined in Eq. (\ref{equation_sample_covariance}).
According to \cite{li2007directional}, CR is an exhaustive method (see Definition \ref{definition_exhaustive_method}) under some mild conditions.

\subsection{Directional Regression (DR)}

We saw in CR method that all the directional information of covariates exists in the set $\{\b{x}_i - \b{x}_j\, |\, 1 \leq i < j \leq n\}$.
Directional Regression (DR), proposed in \cite{li2007directional}, uses this fact. We define:
\begin{align*}
A(Y, Y') := \mathbb{E}[(Z - Z')(Z - Z')^\top\, |\, Y, Y'].
\end{align*}
The directions $(Z - Z')$ which are aligned with $\mathcal{S}_{Y|Z}$ are hugely affected by $Y$ while the directions $(Z - Z')$ aligned with $\mathcal{S}_{Y|Z}^\bot$ are not affected \cite{li2007directional}. 
In DR, we slice the labels into $h$ slices. 

\begin{theorem}[{\citep[Theorem 1 and Section 4]{li2007directional}}]\label{theorem_DR}
With the two conditions mentioned in Section \ref{section_SAVE}, the column space of $2 \b{I} - A(Y, Y')$ is contained in $\mathcal{S}_{Y|Z}$, where $\b{I}$ is the identity matrix. 
Hence, an estimate of $\mathcal{S}_{Y|Z}$ is:
\begin{align*}
\mathbb{E}[2 \b{I} - A(Y, Y')]^2 \approx \frac{1}{\binom{h}{2}} \sum_{k < \ell} (2 \b{I} - \widehat{A}(\mathcal{D}_k, \mathcal{D}_\ell))^2,
\end{align*}
where:
\begin{align*}
&\widehat{A}(\mathcal{D}_k, \mathcal{D}_\ell) \\
&:= \frac{\sum_{i<j} (\widehat{\b{z}}_i - \widehat{\b{z}}_j) (\widehat{\b{z}}_i - \widehat{\b{z}}_j)^\top \mathbb{I}(y_i \in \mathcal{D}_k, y_j \in \mathcal{D}_\ell)}{\sum_{i<j} \mathbb{I}(y_i \in \mathcal{D}_k, y_j \in \mathcal{D}_\ell)},
\end{align*}
is an empirical estimate for $A(Y, Y')$.
Note that $\mathbb{I}(.)$ is the indicator function and $\mathcal{D}_k$ is the $k$-th slice of labels. 
\end{theorem}

We define \cite{li2007directional}:
\begin{align*}
\mathbb{R}^{d \times d} &\ni \widehat{\b{F}} := 2 \sum_{s=1}^h \rho_s E_n^2(\widehat{Z} \widehat{Z}^\top - \b{I}\, |\, Y \in \mathcal{D}_s) \\
&+ 2 \Big(\sum_{s=1}^h \rho_s E_n(\widehat{Z}\, |\, Y \in \mathcal{D}_s) E_n(\widehat{Z}^\top\, |\, Y \in \mathcal{D}_s)\Big)^2 \\
&+ 2 \Big(\sum_{s=1}^h \rho_s E_n(\widehat{Z}^\top\, |\, Y \in \mathcal{D}_s) E_n(\widehat{Z}\, |\, Y \in \mathcal{D}_s) \\
&~~~~\times \sum_{s=1}^h \rho_s E_n(\widehat{Z}\, |\, Y \in \mathcal{D}_s) E_n(\widehat{Z}^\top\, |\, Y \in \mathcal{D}_s) \Big),
\end{align*}
where:
\begin{align*}
E_n(\widehat{Z}\, |\, Y \in \mathcal{D}_s) := \frac{\sum_{i=1}^n \widehat{\b{z}}_i\, \mathbb{I}(y_i \in \mathcal{D}_s)}{\sum_{i=1}^n \mathbb{I}(y_i \in \mathcal{D}_s)}.
\end{align*}
We consider the top $p$ eigenvectors of $\widehat{\b{F}}$ with largest eigenvalues \cite{ghojogh2019eigenvalue}.
According to Theorem \ref{theorem_DR}, these eigenvectors are $\{\b{w}_j\}_{j=1}^p$ used in Eq. (\ref{equation_regression_problem_vector_standardized}). 
We use Eq. (\ref{equation_u_w_relation_estimate}) to calculate $\{\b{u}_j\}_{j=1}^p$ from $\{\b{w}_j\}_{j=1}^p$. The calculated $\{\b{u}_j\}_{j=1}^p$ are the bases for the central subspace and are used in Eq. (\ref{equation_regression_problem_vector}). 
Under some conditions specified in {\citep[Theorem 3]{li2007directional}}, DR is an exhaustive method (see Definition \ref{definition_exhaustive_method}).

\subsection{Likelihood-based Methods}

There are likelihood-based methods in the family of inverse regression methods. They use Maximum Likelihood Estimation (MLE) for estimating the projection matrix onto the central subspace. 
Two fundamental likelihood-based methods are Principal Fitted Components (PFC) \cite{cook2007fisher,cook2008principal} and Likelihood Acquired Direction (LAD) \cite{cook2009likelihood}. There exist some recent likelihood-based methods \cite{bura2015sufficient,bura2016sufficient} which are not covered here for brevity.
Envelope model, proposed in \cite{cook2010envelope} and improved in \cite{zhang2018functional,zhang2020principal}, is also another likelihood-based method not covered here. 

\subsubsection{Principal Fitted Components (PFC)}

Principal Fitted Components (PFC), was proposed in \cite{cook2008principal} and developed over \cite{cook2007fisher}.
It considers the inverse regression problem in Definition \ref{definition_inverse_regression}. 
We assume that the covariates have normal distribution, following Lemma \ref{lemma_linearity_condition}.
We can have \cite{cook2007fisher,cook2008principal}:
\begin{align}\label{equation_PFC_model}
\b{X} | Y = \widehat{\b{\mu}}_x + \b{U} \b{v}_y + \b{\varepsilon},
\end{align}
where $\b{X} \in \mathbb{R}^{d \times n}$ are the covariates, $\b{U} \in \mathbb{R}^{d \times p}$ is the orthogonal projection matrix onto $p$-dimensional central subspace, $\b{v}_y := \b{U}^\top (\mathbb{E}[X|Y] - \widehat{\b{\mu}}_x) \in \mathbb{R}^p$, $\widehat{\b{\mu}}_x$ is defined in Eq. (\ref{equation_sample_mean}), and $\b{\varepsilon}$ is the independent noise. 

\begin{lemma}[\cite{cook2007fisher,cook2008principal}]
In the model of Eq. (\ref{equation_PFC_model}), the dimension reduction:
\begin{align}\label{equation_PFC_projection}
P_\mathcal{S}\b{X} = \b{U}^\top \b{\Sigma}_{\varepsilon\varepsilon}^{-1} \b{X},
\end{align}
is sufficient (see Definition \ref{definition_sufficient_reduction}), where $\b{\Sigma}_{\varepsilon\varepsilon}$ denotes the covariance of noise. 
\end{lemma}

PFC estimates the matrices $\b{U}$ and $\b{\Sigma}_{\varepsilon\varepsilon}$ by MLE to calculate the projection onto the central subspace by Eq. (\ref{equation_PFC_projection}).
The Eq. (\ref{equation_PFC_model}) can be restated as \cite{cook2008principal}:
\begin{align}\label{equation_PFC_model_2}
\b{X} | Y = \b{\mu} + \b{U} \b{\beta} \b{f}(y) + \b{\varepsilon},
\end{align}
where $\b{f}(y)$ is a known vector-valued function of labels. 
We assume that the covariates have normal distribution, following Lemma \ref{lemma_linearity_condition}. 
Hence, the log-likelihood is \cite{cook2008principal}:
\begin{align*}
&L(\b{\mu}, \b{U}, \b{\beta}, \b{\Sigma}_{\varepsilon\varepsilon}) = -\frac{nd}{2} \ln(2\pi) - \frac{n}{2} \ln(|\b{\Sigma}_{\varepsilon\varepsilon}|) \\
&- \frac{1}{2} \sum_{i=1}^n \big(\b{x}_i - \b{\mu} - \b{U} \b{\beta}(\b{f}(y_i) - \bar{\b{f}})^\top\big) \b{\Sigma}_{\varepsilon\varepsilon}^{-1} \\
&~~~~~~~~~~~~~~~~~~\big(\b{x}_i - \b{\mu} - \b{U} \b{\beta}(\b{f}(y_i) - \bar{\b{f}})\big),
\end{align*}
where $\bar{\b{f}}$ is the mean of $\b{f}(y_i)$'s.
Optimizing this log-likelihood iteratively can give us estimations for $\b{U}$ and $\b{\Sigma}_{\varepsilon\varepsilon}$. Using these estimations in Eq. (\ref{equation_PFC_projection}) gives the projection onto the central subspace. 

\subsubsection{Likelihood Acquired Direction (LAD)}

Another likelihood-based method is Likelihood Acquired Direction (LAD) \cite{cook2009likelihood}. 
In LAD, we slice the labels into $h$ slices and use MLE to estimate the central subspace. 

\begin{theorem}[{\citep[Theorem 2]{cook2009likelihood}}]
The MLE of $\mathcal{S}_{Y|X}$ maximizes the log-likelihood:
\begin{align*}
L(\mathcal{S}) = &-\frac{nd}{2} (1+\ln(2\pi)) + \frac{n}{2} \ln(|\b{U}^\top \widehat{\b{\Sigma}}_{xx} \b{U}|_0) \\
&- \frac{n}{2} \ln(|\widehat{\b{\Sigma}}_{\varepsilon\varepsilon}|) - \frac{1}{2} \sum_{s=1}^h \rho_s n \ln(|\b{U}^\top \widehat{\b{\Sigma}}_{\varepsilon\varepsilon} \b{U}|_0),
\end{align*}
over the Grassmannian manifold $\mathcal{G}(p,d)$, where $|.|_0$ denotes the product of the non-zero eigenvalues of matrix. Other notations have been defined before. 
\end{theorem}

In LAD, we use iterative second-order Riemannian optimization \cite{absil2009optimization} on Grassmannian manifold $\mathcal{G}(p,d)$ to find the projection matrix $\b{U} \in \mathbb{R}^{d \times p}$ onto the subspace $\mathcal{S}$.
LAD is an exhaustive dimension reduction method (see Definition \ref{definition_exhaustive_method}) and finds the subspace $\mathcal{S}_{Y|X}$ \cite{cook2009likelihood}.  

\subsection{Graphical Regression}

Graphical regression \cite{cook1998regression,cook1998regressionPaper} uses plots to estimate the central subspace visually. 

\begin{definition}[Sufficient summary plot \cite{cook1998regression}]
Consider projection of covariates onto the subspace by $P_\mathcal{S} \b{X} = \b{U}^\top \b{X}$ where Eq. (\ref{equation_X_Y_independent_projection}) holds. A plot of $\b{y}$ versus $\b{U}^\top \b{x}$ is called the (minimal) minimal sufficient summary plot.
\end{definition}

In this approach, we plot the sufficient summary plot for various $p$ values and the smallest $p$ which shows an acceptable projection in the plot is chosen as the dimensionality of central subspace \cite{cook1998regression}. 
This method is explained in more detail as follows \cite{cook1998regressionPaper}. 
Let $x^j$ be denoted by the $j$-th dimension and let $\b{x}^{\setminus j_1, j_2}$ denote all the dimensions of $\b{x}$ except $x^{j_1}$ and $x^{j_2}$.
We start from projecting the first two dimensions onto a subspace by and check visually if this holds \cite{cook1998regressionPaper}:
\begin{align*}
X \ind Y\, |\, (c_1 x^1 + c_2 x^2, \b{x}^{\setminus 1,2}).
\end{align*}
If this holds, we combine the two dimensions $x^1$ and $x^2$ into one new dimension, denoted by $x^{1,2}$. This reduces the dimension of covariate by one by projecting it onto a $(d-1)$-dimensional subspace. We repeat this for other dimensions until the conditional independence does not hold anymore visually. 
Some more explanation on graphical regressions can be found in \cite{cook1994interpretation,cook2009introduction}.
Also, see \cite{cook1996graphics} for graphical regression on binary labels.

\section{Forward Regression Methods}\label{section_forward_regression_methods}

Forward regression methods are not based on inverse regression. Some important forward regression methods are Principal Hessian Directions (pHd) \cite{li1992principal}, Minimum Average Variance Estimation (MAVE) \cite{xia2002adaptive}, Conditional Variance Estimation (CVE) \cite{fertl2021conditional}, and deep SDR \cite{banijamali2018deep,kapla2021fusing}. In the following, we introduce these methods. 

\subsection{Principal Hessian Directions (pHd)}

The method Principal Hessian Directions (pHd), proposed in \cite{li1992principal}, uses Hessian matrix and the Stein's lemma to estimate the central subspace. 
Consider the Hessian matrix of the covariates denoted by $\b{H}_x$. Let the mean of Hessian matrix be $\bar{\b{H}}_x$. 

\begin{lemma}[{\citep[Section 2.1]{li1992principal}}]
The top $p$ eigenvectors of $\bar{\b{H}}_x \b{\Sigma}_{xx}$, with largest eigenvalues, are the bases of the $p$-dimensional central subspace with projection matrix $\b{U} \in \mathbb{R}^{d \times p}$.
\end{lemma}

\begin{lemma}[Stein's lemma {\citep[Lemma 4]{stein1981estimation}}]
Let $g(.)$ be a function and a random variable $X$ have a mean $\b{\mu}$ and variance one. We have:
\begin{align*}
& \mathbb{E}[X - \b{\mu}]\, g(X) = \mathbb{E}\big[\frac{\partial g(X)}{\partial X}\big] \\
& \mathbb{E}[X - \b{\mu}]^2\, g(X) = \mathbb{E}[g(X)] + \mathbb{E}\big[\frac{\partial^2 g(X)}{\partial X^2}\big].
\end{align*}
\end{lemma}

\begin{theorem}[{\citep[Lemma 3.1, Corollary 3.1, and Theorem 3.1]{li1992principal}}]
Let:
\begin{align*}
\mathbb{R}^{d \times d} \ni \b{\Sigma}_{yxx} = \mathbb{E}[(Y - \mu_y) (X - \b{\mu}_x) (X - \b{\mu}_x)^\top],
\end{align*}
where $\b{\mu}_x$ and $\mu_y$ are the mean of covariates and labels, respectively. 
Based on the Stien's lemma, we have:
\begin{align*}
\bar{\b{H}}_x = \b{\Sigma}_{xx}^{-1} \b{\Sigma}_{yxx} \b{\Sigma}_{xx}^{-1}.
\end{align*}
If the covariates have normal distribution, the bases of central subspace are the top $p$ eigenvectors of the following generalized eigenvalue problem:
\begin{align*}
\b{\Sigma}_{yxx} \b{u}_j = \lambda_j \b{\Sigma}_{xx} \b{u}_j, \quad \forall j \in \{1, \dots, p\},
\end{align*}
with largest eigenvalues, where $\b{u}_j$'s and $\lambda_j$'s are the eigenvectors and eigenvalues, respectively \cite{ghojogh2019eigenvalue}. 
\end{theorem}

We can estimate the covariance matrix $\b{\Sigma}_{yxx}$ by:
\begin{align*}
\mathbb{R}^{d \times d} \ni \widehat{\b{\Sigma}}_{yxx} = \frac{1}{n} \sum_{i=1}^n (y_i - \widehat{\mu}_y) (\b{x}_i - \widehat{\b{\mu}}_x) (\b{x}_i - \widehat{\b{\mu}}_x)^\top,
\end{align*}
where $\widehat{\b{\mu}}_x$ and $\widehat{\mu}_y$ are the sample mean of covariates and labels, respectively. 
Hence, we can find the eigenvectors of the generalized eigenvalue problem $\widehat{\b{\Sigma}}_{yxx} \b{u}_j = \lambda_j \widehat{\b{\Sigma}}_{xx} \b{u}_j, \forall j$ to have the projection matrix $\b{U} = [\b{u}_1, \dots, \b{u}_p]$ onto the central subspace. 
According to \cite{li2007directional}, pHd is not an exhaustive method (see Definition \ref{definition_exhaustive_method}). 
The pHd method has been further improved in \cite{cook1998principal,cook2002dimension,cook2004determining}. Some technical comments on it are available in \cite{li1998principal}.

\subsection{Minimum Average Variance Estimation (MAVE)}

Minimum Average Variance Estimation (MAVE) was first proposed in \cite{xia2002adaptive}.

\begin{theorem}[{\citep[Section 2]{xia2002adaptive}}]
The orthogonal projection matrix onto the central subspace is the solution to the following optimization problem:
\begin{equation*}
\begin{aligned}
& \underset{\b{U}}{\text{minimize}}
& & \mathbb{E}\big[Y - \mathbb{E}[Y\, |\, \b{U}^\top X]\big]^2 \\
& \text{subject to}
& & \b{U}^\top \b{U} = \b{I}.
\end{aligned}
\end{equation*}
We define $\sigma^2_{\b{U}} (\b{U}^\top X) := \mathbb{E}\big[(Y - \mathbb{E}[Y\, |\, \b{U}^\top X])^2 | \b{U}^\top X\big]$. We have:
\begin{align*}
\mathbb{E}\big[Y - \mathbb{E}[Y\, |\, \b{U}^\top X]\big]^2 = \mathbb{E}[\sigma^2_{\b{U}} (\b{U}^\top X)].
\end{align*}
Hence, we have:
\begin{equation}\label{equation_MAVE_optimization}
\begin{aligned}
& \underset{\b{U}}{\text{minimize}}
& & \mathbb{E}[\sigma^2_{\b{U}} (\b{U}^\top X)] \\
& \text{subject to}
& & \b{U}^\top \b{U} = \b{I}.
\end{aligned}
\end{equation}
\end{theorem}

We can estimate $\sigma^2_{\b{U}} (\b{U}^\top X)$ by {\citep[Section 2]{xia2002adaptive}}:
\begin{align*}
&\widehat{\sigma}^2_{\b{U}} (\b{U}^\top \b{x}_j) = \\
&\min_{a \in \mathbb{R}, \b{b} \in \mathbb{R}^p} \sum_{i=1}^n \Big(y_i - \big(a + \b{b}^\top \b{U}^\top (\b{x}_i - \b{x}_j)\big)\Big)^2 w_{ij},
\end{align*}
where:
\begin{align*}
w_{ij} := \frac{k_h(\b{U}^\top (\b{x}_i - \b{x}_j))}{\sum_{\ell=1}^n k_h(\b{U}^\top (\b{x}_\ell - \b{x}_j))},
\end{align*}
in which $k_h(.)$ denotes a kernel function with bandwidth $h$ \cite{ghojogh2021reproducing}. 
We can use the estimated $\widehat{\sigma}^2_{\b{U}} (\b{U}^\top \b{x}_j)$ in Eq. (\ref{equation_MAVE_optimization}) to calculate the projection matrix onto the central subspace. The optimization variables of this problem are $a$, $\b{b}$, and $\b{U}$ and any optimization method can be used in an alternative optimization approach for solving this problem \cite{ghojogh2021kkt}.

The MAVE method has been improved in \cite{xia2007constructive}. Some other methods based on MAVE are central subspace MAVE (csMAVE) \cite{wang2008sliced} and ensemble MAVE \cite{yin2011sufficient2}.
An R programming language package for MAVE exists \cite{weiqiang2019mave}.


\subsection{Conditional Variance Estimation (CVE)}

Conditional Variance Estimation (CVE) was first proposed in \cite{fertl2021conditional,fertl2021sufficient}. 

\begin{theorem}[{\citep[Section 2, Section 2.1, Coroallry 2]{fertl2021conditional}}]
We define:
\begin{align*}
& \widetilde{L}(\b{V}, \b{s}_0) := \mathbb{V}\text{ar}(Y | X \in \b{s}_0 + \mathbb{C}\text{ol}(\b{V})), \\
& L(\b{V}) := \mathbb{E}[\widetilde{L}(\b{V}, X)],
\end{align*}
where $\b{s}_0 \in \mathbb{R}^d$ is a shifting point and $\mathbb{C}\text{ol}(\b{V})$ is column space of $\b{V}$.
Consider the following optimization problem:
\begin{equation}\label{equation_CVE_optimization}
\begin{aligned}
& \underset{\b{V}}{\text{minimize}}
& & L(\b{V}) \\
& \text{subject to}
& & \b{V} \in \mathcal{S}t(p,d),
\end{aligned}
\end{equation}
whose solution we denote by $\b{V}_p$.
The orthogonal space to $\mathbb{C}\text{ol}(\b{V}_p)$ is equal to the columns space of the projection matrix $\b{U}$ onto the central space:
\begin{align}\label{equation_CVE_column_space_U}
\mathbb{C}\text{ol}(\b{U}) = \mathbb{C}\text{ol}(\b{V}_p)^{\bot}.
\end{align}
\end{theorem}

We can estimate $L(\b{V})$ as follows. We define \cite{fertl2021conditional}:
\begin{align*}
& d_i(\b{V}, \b{s}_0) := \|\b{U}^\top (\b{x}_i - \b{s}_0)\|_2^2, \\
& w_i(\b{V}, \b{s}_0) := \frac{k_h(d_i(\b{V}, \b{s}_0))}{\sum_{\ell=1}^n k_h(d_\ell(\b{V}, \b{s}_0))}, 
\end{align*}
where $w_i$ is defined similar to the $w_i$ in MAVE method. 
The estimate of $\widetilde{L}(\b{V}, \b{s}_0)$ is \cite{fertl2021conditional}:
\begin{align*}
\widetilde{L}_n(\b{V}, \b{s}_0) = \sum_{i=1}^n w_i(\b{V}, \b{s}_0) y_i^2 - \Big(\sum_{i=1}^n w_i(\b{V}, \b{s}_0) y_i\Big)^2.
\end{align*}
Finally, the estimate of $L(\b{V})$ is:
\begin{align*}
L_n(\b{V}) = \frac{1}{n} \sum_{j=1}^n \widetilde{L}_n(\b{V}, \b{x}_j). 
\end{align*}
We use this $L_n(\b{V})$ in Eq. (\ref{equation_CVE_optimization}) instead of $L(\b{V})$ to find the matrix $\b{V}$ using Riemannian optimization on Stiefel manifold \cite{absil2009optimization}. 
Then, using Eq. (\ref{equation_CVE_column_space_U}), we find the projection matrix $\b{U}$ onto the central subspace. 
It is noteworthy that there also exists an ensemble version of CVE \cite{fertl2021ensemble}.

\subsection{Deep Sufficient Dimension Reduction}\label{section_deep_SDR}

\subsubsection{Deep Variational Sufficient Dimension Reduction (DVSDR)}

Deep Variational Sufficient Dimension Reduction (DVSDR), proposed in \cite{banijamali2018deep}, is one of the approaches for deep SDR. 
It formulates the network as a variational autoencoder \cite{kingma2014auto,ghojogh2021factor} and interprets it as SDR. 
Let $\mathbb{R}^{p \times n} \ni \widetilde{\b{X}} = \b{U}^\top \b{X}$ be the projected covariates onto the central subspace. 
According to Eq. (\ref{equation_X_Y_independent_projection_2}), we have $X \ind Y\, |\, \widetilde{X}$ where $\widetilde{X}$ is the random variable associated to $\widetilde{\b{X}}$. Hence, we can see $\widetilde{X}$ as a latent factor on which both covariates $X$ and labels $Y$ depend but conditioning on that, they are independent. Recall that in variational inference, we also have a latent factor (see \cite{ghojogh2021factor}). 
As in variational autoencoder, we can have a deep autoencoder whose encoder and decoder model the conditional probabilities $q(\widetilde{X} | X)$ and $p(X | \widetilde{X})$, respectively. 
Let the weights of encoder and decoder be denoted by $\phi$ and $\theta$, respectively. 
The encoder layers map data from $d$ to $p$ dimensions and the decoder layers map data from $p$ to $d$ dimensions.
We can use the Evidence Lower Bound (ELBO) of variational inference as \cite{ghojogh2021factor}:
\begin{align*}
\mathcal{L}_{\phi, \theta}^u(\b{x}_i) := &\,\mathbb{E}_{q_{\phi}(\widetilde{X} | X)}\big[\ln(p_\theta(\b{x}_i | \widetilde{\b{x}}_i))\big] \\
&- \text{KL}\big(q_\phi(\widetilde{\b{x}}_i | \b{x}_i)\, \|\, p(\widetilde{\b{x}}_i) \big).
\end{align*}
We can also add additional layers in the decoder part to generate labels for classification or regression. Let the weights of these layers be denoted by $\psi$. These layers model $p(Y | \widetilde{X})$. Including these layers into ELBO gives:
\begin{align*}
&\mathcal{L}_{\phi, \theta, \psi}^\ell(\b{x}_i) := \mathbb{E}_{q_{\phi}(\widetilde{X} | X)}\big[\ln(p_\theta(\b{x}_i | \widetilde{\b{x}}_i))\big] \\
&+ \mathbb{E}_{q_{\phi}(\widetilde{X} | X)}\big[\ln(p_\psi(y_i | \widetilde{\b{x}}_i))\big] - \text{KL}\big(q_\phi(\widetilde{\b{x}}_i | \b{x}_i)\, \|\, p(\widetilde{\b{x}}_i) \big).
\end{align*}
We can make a regularized loss function using both of these ELBO loss functions:
\begin{align*}
\max_{\phi, \theta, \psi} \sum_{i=1}^n \big( \mathcal{L}_{\phi, \theta, \psi}^\ell(\b{x}_i) + \mathcal{L}_{\phi, \theta}^u(\b{x}_i) \big). 
\end{align*}
The weights $\phi, \theta, \psi$ are tuned by backpropagation. 
The reader can refer to \cite{ghojogh2021factor} to learn more on how to train a variational autoencoder. 
If the dataset is partially labeled, we can also use $\mathcal{L}_{\phi, \theta, \psi}^\ell$ and $\mathcal{L}_{\phi, \theta}^u$ for labeled and unlabeled parts, respectively, to have semi-supervised learning.  

\subsubsection{Meta-Learning for Sufficient Dimension Reduction}

Another approach for deep SDR is proposed in \cite{kapla2021fusing}. This approach can be seen as meta learning \cite{finn2017model} although the authors of that paper do not mention meta learning in their paper. 
Consider the model of Eq. (\ref{equation_regression_problem_vector_2}) for regression. 
This can be modeled by a neural network whose first layer projects data $\b{X}$ from $d$ dimensions to $p$ dimensions. The weights of this layer is the matrix $\b{U} \in \mathbb{R}^{d \times p}$. 
The last layer has one neuron for predicting the label $y$. 
Let the weights of network be denoted by $\theta$. The first layer models $f(\b{U}^\top \b{X})$ in (\ref{equation_regression_problem_vector_2}). First, we feed the covariates $\b{X}$ to the network with weights $\theta$, obtain some predicted labels, and train the weights using backpropagation with least squares loss between the labels and predicted labels. Let $\theta'$ be the updated weights by this training. Then, in the meta-training phase, the covariates are fed to the network with the weights $\theta'$, obtain some predicted labels, and train the weights using backpropagation with least squares loss between the labels and predicted labels. This procedure of training and meta-training is repeated until convergence. 

\section{Kernel Dimension Reduction}\label{section_KDR_methods}

\subsection{Supervised Kernel Dimension Reduction}

Kernel Dimension Reduction (KDR) was originally proposed in \cite{fukumizu2003kernel} and further discussed in \cite{fukumizu2004dimensionality,fukumizu2009kernel}.
It is one of the methods for SDR and it makes use of Eqs. (\ref{equation_effective_subspace_1}) and (\ref{equation_effective_subspace_2}) for finding the sufficient subspace. 
Its approach is more toward machine learning for dimensionality reduction \cite{fukumizu2003kernel,fukumizu2004dimensionality}; although, it can also be considered a statistical method for high-dimensional regression \cite{fukumizu2009kernel}.
KDR uses kernels in the Reproducing Kernel Hilbert Space (RKHS) \cite{ghojogh2021reproducing} for calculating the central subspace. 
It is noteworthy that another SDR method which uses RKHS is kernel Principal Support Vector Machines (PSVM) \cite{li2011principal}. 

\begin{definition}[Conditional covariance operator \cite{fukumizu2003kernel}]
The conditional covariance operator in RKHS is defined as:
\begin{align}
\b{\Sigma}_{YY|\widetilde{X}} := \b{\Sigma}_{YY} - \b{\Sigma}_{Y\widetilde{X}} \b{\Sigma}_{\widetilde{X}\widetilde{X}}^{-1} \b{\Sigma}_{\widetilde{X}Y}
\end{align}
where $\b{\Sigma}_{AB}$ is the covariance of random variables $A$ and $B$ and $\widetilde{\b{X}} := \b{U}^\top \b{X}$ is the projected covariates ($\widetilde{X}$ is its associated random variable). 
\end{definition}

\begin{theorem}[{\citep[Theorem 5]{fukumizu2003kernel}}]
Let $\mathbb{R}^{d \times d} \ni \b{Q} = [\b{U} | \b{V}]$ be an orthogonal matrix, where $\b{U} \in \mathbb{R}^{d \times p}$ is the truncated projection matrix onto the $p$-dimensional subspace and $\b{V} \in \mathbb{R}^{d \times (d-p)}$ is the rest of matrix $\b{Q}$.
We have:
\begin{align*}
&\b{\Sigma}_{YY|\widetilde{X}} \geq \b{\Sigma}_{YY|X}, \\
& \b{\Sigma}_{YY|\widetilde{X}} = \b{\Sigma}_{YY|X} \iff Y \ind (\b{V}^\top X)\, |\, \widetilde{X},
\end{align*}
where $\widetilde{\b{X}} := \b{U}^\top \b{X}$ is the projected covariates onto the $p$-dimensional central subspace.
The central subspace can be found by minimizing the covariance of labels conditioned on the projected covariates onto the subspace:
\begin{equation}\label{equation_KDR_optimization}
\begin{aligned}
& \underset{\b{U}}{\text{minimize}} 
& & \b{\Sigma}_{YY|\widetilde{X}}.
\end{aligned}
\end{equation}
\end{theorem}

KDR allows the labels to be multi-dimensional ($\ell$-dimensional), i.e., $\b{Y} \in \mathbb{R}^{\ell \times n}$.
Consider the double-centered kernel matrices for labels and the projected covariates:
\begin{align}
&\mathbb{R}^{n \times n} \ni \widehat{\b{K}}_Y = \b{H} \b{Y}^\top \b{Y} \b{H}, \\
&\mathbb{R}^{n \times n} \ni \widehat{\b{K}}_{\widetilde{X}} = \b{H} \widetilde{\b{X}}^\top \widetilde{\b{X}} \b{H} = \b{H} \b{X}^\top \b{U} \b{U}^\top \b{X} \b{H}, \label{equation_KDR_kernel_projected_X_centered}
\end{align}
where $\b{H} := \b{I} - (1/n) \b{1}\b{1}^\top \in \mathbb{R}^{n \times n}$ is the centering matrix (see \cite{ghojogh2019unsupervised,ghojogh2021reproducing} for more details on the centering matrix).
The empirical estimates for the covariance matrices are \cite{fukumizu2003kernel}:
\begin{align*}
& \widehat{\b{\Sigma}}_{YY} = (\widehat{\b{K}}_Y + \epsilon \b{I})^2, \quad \widehat{\b{\Sigma}}_{Y\widetilde{X}} = \widehat{\b{K}}_Y \widehat{\b{K}}_{\widetilde{X}}, \\
& \widehat{\b{\Sigma}}_{UU} = (\widehat{\b{K}}_U + \epsilon \b{I})^2, \quad \widehat{\b{\Sigma}}_{\widetilde{X}Y} = \widehat{\b{K}}_{\widetilde{X}} \widehat{\b{K}}_Y,
\end{align*}
where $\epsilon$ is a small positive number and adding $\epsilon \b{I}$ is to make the kernel matrices full rank and invertible. 
The $\b{\Sigma}_{YY|\widetilde{X}} = \b{\Sigma}_{YY|\b{U}^\top X}$ can be estimated empirically as \cite{fukumizu2003kernel}:
\begin{equation}\label{equation_KDR_Cov_YY_X_tilde}
\begin{aligned}
&\mathbb{R}^{n \times n} \ni \widehat{\b{\Sigma}}_{YY|\widetilde{X}} = \widehat{\b{\Sigma}}_{YY} - \widehat{\b{\Sigma}}_{Y\widetilde{X}} \widehat{\b{\Sigma}}_{\widetilde{X}\widetilde{X}}^{-1} \widehat{\b{\Sigma}}_{\widetilde{X}Y} \\
&= (\widehat{\b{K}}_Y + \epsilon \b{I})^2 - \widehat{\b{K}}_Y \widehat{\b{K}}_{\widetilde{X}} (\widehat{\b{K}}_{\widetilde{X}} + \epsilon \b{I})^{-2} \widehat{\b{K}}_{\widetilde{X}} \widehat{\b{K}}_Y.
\end{aligned}
\end{equation}

\subsubsection{Supervised KDR by Projected Gradient Descent}

In practice, we can use the determinant of estimated covariance matrix, i.e., $\text{det}(\widehat{\b{\Sigma}}_{YY|\widetilde{X}})$ in Eq. (\ref{equation_KDR_optimization}). 
According to Schur complement, we have \cite{fukumizu2003kernel}:
\begin{align}\label{equation_KDR_det_frac}
\text{det}(\widehat{\b{\Sigma}}_{YY|\widetilde{X}}) = \frac{\text{det}(\widehat{\b{\Sigma}}_{(Y\widetilde{X})(Y\widetilde{X})})}{\text{det}(\widehat{\b{\Sigma}}_{\widetilde{X}\widetilde{X}})},
\end{align}
where:
\begin{align*}
\widehat{\b{\Sigma}}_{(Y\widetilde{X})(Y\widetilde{X})} &= 
\begin{bmatrix}
\widehat{\b{\Sigma}}_{YY} & \widehat{\b{\Sigma}}_{Y\widetilde{X}} \\
\widehat{\b{\Sigma}}_{\widetilde{X}Y} & \widehat{\b{\Sigma}}_{\widetilde{X}\widetilde{X}}
\end{bmatrix}.
\end{align*}
We symmetrize Eq. (\ref{equation_KDR_det_frac}) by dividing it by the constant $\text{det}(\widehat{\b{\Sigma}}_{YY})$. Finally, in practice, Eq. (\ref{equation_KDR_optimization}) is stated as:
\begin{equation}\label{equation_KDR_optimization_practice}
\begin{aligned}
& \underset{\b{U}}{\text{minimize}} 
& & \frac{\text{det}(\widehat{\b{\Sigma}}_{(Y\widetilde{X})(Y\widetilde{X})})}{\text{det}(\widehat{\b{\Sigma}}_{YY})\, \text{det}(\widehat{\b{\Sigma}}_{\widetilde{X}\widetilde{X}})}.
\end{aligned}
\end{equation}
We can solve this problem iteratively by gradient descent \cite{ghojogh2021kkt} where the gradient is \cite{fukumizu2004dimensionality}:
\begin{align*}
&\frac{\partial \ln (\text{det}(\widehat{\b{\Sigma}}_{YY|\widetilde{X}}))}{\partial \b{U}} = \textbf{tr}(\widehat{\b{\Sigma}}_{YY|\widetilde{X}} \frac{\partial \widehat{\b{\Sigma}}_{YY|\widetilde{X}}}{\partial \b{U}}) \\
&= 2\epsilon\, \textbf{tr}\Big(\widehat{\b{\Sigma}}_{YY|\widetilde{X}}^{-1} \widehat{\b{K}}_Y (\widehat{\b{K}}_{\widetilde{X}} + \epsilon \b{I})^{-1} \frac{\partial \widehat{\b{K}}_{\widetilde{X}}}{\partial \b{U}} \\
&~~~~~~~~~~~~~~~~~~~~~~~~~~~~~~~~~~~~~~(\widehat{\b{K}}_{\widetilde{X}} + \epsilon \b{I})^{-2} \widehat{\b{K}}_{\widetilde{X}} \widehat{\b{K}}_Y\Big).
\end{align*}

\subsubsection{Supervised KDR by Riemannian Optimization}

Another approach for KDR optimization is as follows \cite{fukumizu2009kernel}. We use Eq. (\ref{equation_KDR_Cov_YY_X_tilde}) in Eq. (\ref{equation_KDR_optimization}) where we strengthen the diagonal of $\widehat{\b{\Sigma}}_{\widetilde{X}\widetilde{X}}$ to make it invertible. Paper \cite{fukumizu2009kernel} makes optimization constrained by putting constraint on the projection matrix to be orthogonal, i.e., $\b{U}^\top \b{U} = \b{I}$. According to Definition \ref{definition_Stiefel_Grassmannian_manifolds}, this constraint means that the projection matrix belongs to the Stiefel manifold. Hence, the optimization problem is:
\begin{equation}\label{equation_KDR_optimization_Stiefel}
\begin{aligned}
& \underset{\b{U}}{\text{minimize}} 
& & \widehat{\b{\Sigma}}_{YY} - \widehat{\b{\Sigma}}_{Y\widetilde{X}} (\widehat{\b{\Sigma}}_{\widetilde{X}\widetilde{X}} + \epsilon \b{I})^{-1} \widehat{\b{\Sigma}}_{\widetilde{X}Y} \\
& \text{subject to}
& & \b{U} \in \mathcal{S}t(p,d),
\end{aligned}
\end{equation}
which can be solved iteratively by Riemannian optimization \cite{absil2009optimization}.

The Eq. (\ref{equation_KDR_optimization_Stiefel}) can be slightly changed and restated to \cite{nilsson2007regression}:
\begin{equation}\label{equation_KDR_optimization_Ky_Kx}
\begin{aligned}
& \underset{\b{U}}{\text{minimize}} 
& & \textbf{tr}\big(\widehat{\b{K}}_Y (\widehat{\b{K}}_{\widetilde{X}} + n\epsilon \b{I})^{-1}\big) \\
& \text{subject to}
& & \b{U}^\top \b{U} = \b{I}.
\end{aligned}
\end{equation}
The Eq. (\ref{equation_KDR_optimization_Ky_Kx}) is also used as the optimization problem for KDR.

\subsubsection{Formulation of Supervised KDR by HSIC}

Suppose we want to measure the dependence of two random variables. Measuring the correlation between them is easier because correlation is just ``linear'' dependence. 
According to \cite{hein2004kernels}, two random variables are independent if and only if any bounded continuous functions of them are uncorrelated. Therefore, if we map the two random variables $\b{x}_1$ and $\b{x}_1$ to two different (``separable'') RKHSs and have $\b{\phi}(\b{x}_1)$ and $\b{\phi}(\b{x}_2)$, we can measure the correlation of $\b{\phi}(\b{x}_1)$ and $\b{\phi}(\b{x}_2)$ in the Hilbert space to have an estimation of dependence of $\b{x}$ and $\b{y}$ in the original space. 
The correlation of $\b{\phi}(\b{x}_1)$ and $\b{\phi}(\b{x}_2)$ can be computed by the Hilbert-Schmidt norm of the cross-covariance of them. 
An empirical not-normalized estimation of the HSIC is introduced \cite{gretton2005measuring}:
\begin{align}\label{equation_HSIC}
\text{HSIC}(\b{x}_1, \b{x}_2) := \textbf{tr}(\b{H}\b{K}_{x_1}\b{H}\b{K}_{x_2}),
\end{align}
where $\b{K}_{x_1}$ and $\b{K}_{x_2}$ are the kernels over $\b{x}_1$ and $\b{x}_2$, respectively.

\begin{theorem}[{\citep[Proposition 1]{wang2010unsupervised}}]
Let $c_0$ be a positive constant and $\epsilon_n^2 \rightarrow 0$ as $n \rightarrow \infty$. 
We have:
\begin{align}
\textbf{tr}\big(\widehat{\b{K}}_Y (\widehat{\b{K}}_{\widetilde{X}} + n\epsilon \b{I})^{-1}\big) &\approx - c_0 n^2 \epsilon_n^2 \textbf{tr}(\b{H}\b{K}_{\widetilde{X}}\b{H}\b{K}_{Y}) \nonumber\\
&= - c_0 n^2 \epsilon_n^2 \textbf{tr}(\widehat{\b{K}}_{\widetilde{X}}\b{K}_{Y}), \label{equation_KDR_HSIC_theorem}
\end{align}
where $\b{K}_{Y} := \b{Y}^\top \b{Y}$.
\end{theorem}
Comparing Eqs. (\ref{equation_KDR_optimization_Ky_Kx}), (\ref{equation_HSIC}), and (\ref{equation_KDR_HSIC_theorem}) shows that the optimization of KDR can also be stated as the following maximization problem:
\begin{equation}\label{equation_KDR_optimization_HSIC}
\begin{aligned}
& \underset{\b{U}}{\text{maximize}} 
& & \text{HSIC}(\widetilde{\b{X}}, \b{Y}) = \textbf{tr}(\widehat{\b{K}}_{\widetilde{X}}\b{K}_{Y}) \\
& \text{subject to}
& & \b{U}^\top \b{U} = \b{I}.
\end{aligned}
\end{equation}

\begin{corollary}[Equivalency of supervised KDR and supervised PCA]
The optimization problem of supervised Principal Component Analysis (PCA) \cite{barshan2011supervised} is exactly the same as Eq. (\ref{equation_KDR_optimization_HSIC}) (note that in HSIC, any of the two kernels can be double-centered and it is not important which kernel is centered but one of them must be double-centered). This equation is the optimization of supervised PCA because it can be shown that PCA is a special case of this problem where the information of labels is not used \cite{ghojogh2019unsupervised}. Hence, supervised KDR and supervised PCA are equivalent!
\end{corollary}

\subsection{Supervised KDR for Nonlinear Regression}

Manifold KDR (mKDR) \cite{nilsson2007regression} performs KDR on manifolds for nonlinear regression.  
mKDR combines the ideas of KDR and Laplacian eigenmap \cite{belkin2001laplacian,ghojogh2021laplacian}. The Laplacian eigenmap brings the nonlinear information of manifold of data into the formulation. 

Let $\{\b{r}_j \in \mathbb{R}^n\}_{j=1}^m$ be the $m$ tailing eigenvectors of the Laplacian matrix of the graph of covariates (n.b. we ignore the eigenvector with eigenvalue zero). We denote $\{\b{t}_i \in \mathbb{R}^m\}_{i=1}^n$ to be $\b{T} := [\b{t}_1, \dots, \b{t}_n] = [\b{r}_1, \dots, \b{r}_m]^\top \in \mathbb{R}^{m \times n}$.
In mKDR, we use kernelization by representation theory \cite{ghojogh2021reproducing} to kernelize the above optimization problem. According to the representation theory, if we pull the centered projected covariates $\b{X} \b{H}$ to RKHS, they must lie in the span of all pulled training points \cite{ghojogh2021reproducing}:
\begin{align*}
&\widetilde{\b{X}} \b{H} = \b{\Phi}(\b{X}) \b{T} \\
&\implies 
\widehat{\b{K}}_{\widetilde{X}} \overset{(\ref{equation_KDR_kernel_projected_X_centered})}{=} \b{T}^\top \b{\Phi}(\b{X})^\top \b{\Phi}(\b{X}) \b{T} = \b{T}^\top \b{K}_x \b{T},
\end{align*}
where $\b{\Phi}(\b{X})$ is the pull of $\b{X}$ to the RKHS and we define the kernel \cite{ghojogh2021reproducing}:
\begin{align}\label{equation_KDR_kernel_Kx}
\b{K}_x := \b{\Phi}(\b{X})^\top \b{\Phi}(\b{X}),
\end{align}
which must be positive semi-definite according to the properties of kernel \cite{ghojogh2021reproducing}. For making the kernel bounded, we also set $\textbf{tr}(\b{K}_x) = 1$.
Substituting all these in Eq. (\ref{equation_KDR_optimization_Ky_Kx}) gives:
\begin{equation}\label{equation_KDR_optimization_Ky_Kx_2}
\begin{aligned}
& \underset{\b{K}_x}{\text{minimize}} 
& & \textbf{tr}\big(\widehat{\b{K}}_Y (\b{T}^\top \b{K}_x \b{T} + n \epsilon \b{I})^{-1}\big) \\
& \text{subject to}
& & \b{K}_x \succeq \b{0}, \\
& & & \textbf{tr}(\b{K}_x) = 1.
\end{aligned}
\end{equation}
Note that, as was explained before, $\b{T}$ is obtained from the Laplacian eigenmap and then used in above optimization. 
We solve Eq. (\ref{equation_KDR_optimization_Ky_Kx_2}) using projected gradient method where every step is projected onto the positive semi-definite cone \cite{ghojogh2021kkt}. After solving the optimization problem, the solution $\b{K}_x$ is decomposed using eigenvalue decomposition (which can be done because it is positive semi-definite so its eigenvalues are not negative):
\begin{align*}
\b{K}_x = \b{A} \b{\Delta} \b{A}^\top = \b{A} \b{\Delta}^{1/2} \b{\Delta}^{1/2} \b{A}^\top \overset{(\ref{equation_KDR_kernel_Kx})}{=} \b{\Phi}(\b{X})^\top \b{\Phi}(\b{X}),
\end{align*}
where $\b{A}$ is a matrix whose columns are the eigenvectors and $\b{\Delta}$ is the diagonal matrix of eigenvalues. 
Therefore, from the above expression, we have $\b{\Phi}(\b{X}) = \b{\Delta}^{1/2} \b{A}^\top$. If we truncate this matrix $\b{\Phi}(\b{X})$ to have the $p$ top eigenvectors with largest eigenvalues, it is the $p$-dimensional embedding of covariates into the subspace. 

\subsection{Unsupervised Kernel Dimension Reduction}

Unsupervised KDR \cite{wang2010unsupervised} does not use labels and can be used when labels are not available. 
Let $X'$ be a copy of random variable $X$. In unsupervised KDR, rather than Eq. (\ref{equation_X_Y_independent_projection_2}), we consider:
\begin{align}\label{equation_X_Y_independent_projection_2_unsupervised}
X \ind X'\, |\, \b{U}^\top X,
\end{align}
meaning that projection onto the central subspace is sufficient for the covariates to be independent of each other. 
We use Eq. (\ref{equation_KDR_optimization_Ky_Kx}) but with $X$ instead of $Y$:
\begin{equation}\label{equation_KDR_optimization_unsupervised}
\begin{aligned}
& \underset{\b{U}}{\text{minimize}} 
& & \textbf{tr}\big(\widehat{\b{K}}_X (\widehat{\b{K}}_{\widetilde{X}} + n\epsilon \b{I})^{-1}\big) \\
& \text{subject to}
& & \b{U}^\top \b{U} = \b{I}.
\end{aligned}
\end{equation}
Any numerical optimization method can be used for solving this problem to find the projection matrix $\b{U}$ onto the central subspace. 










\section{Conclusion}\label{section_conclusion}

In this paper, we introduced and explained different SDR methods including inverse regression methods, forward regression methods, and KDR methods. We showed the SDR methods can be used for both high-dimensional regression and low-dimensional embedding in statistics and machine learning perspectives, respectively.




\bibliography{References}

\begin{thebibliography}{75}
\providecommand{\natexlab}[1]{#1}
\providecommand{\url}[1]{\texttt{#1}}
\expandafter\ifx\csname urlstyle\endcsname\relax
  \providecommand{\doi}[1]{doi: #1}\else
  \providecommand{\doi}{doi: \begingroup \urlstyle{rm}\Url}\fi

\bibitem[Absil et~al.(2009)Absil, Mahony, and Sepulchre]{absil2009optimization}
Absil, P-A, Mahony, Robert, and Sepulchre, Rodolphe.
\newblock \emph{Optimization algorithms on matrix manifolds}.
\newblock Princeton University Press, 2009.

\bibitem[Adragni \& Cook(2009)Adragni and Cook]{adragni2009sufficient}
Adragni, Kofi~P and Cook, R~Dennis.
\newblock Sufficient dimension reduction and prediction in regression.
\newblock \emph{Philosophical Transactions of the Royal Society A:
  Mathematical, Physical and Engineering Sciences}, 367\penalty0
  (1906):\penalty0 4385--4405, 2009.

\bibitem[Banijamali et~al.(2018)Banijamali, Karimi, and
  Ghodsi]{banijamali2018deep}
Banijamali, Ershad, Karimi, Amir-Hossein, and Ghodsi, Ali.
\newblock Deep variational sufficient dimensionality reduction.
\newblock In \emph{Third workshop on Bayesian Deep Learning (NeurIPS 2018)},
  2018.

\bibitem[Barshan et~al.(2011)Barshan, Ghodsi, Azimifar, and
  Jahromi]{barshan2011supervised}
Barshan, Elnaz, Ghodsi, Ali, Azimifar, Zohreh, and Jahromi, Mansoor~Zolghadri.
\newblock Supervised principal component analysis: Visualization,
  classification and regression on subspaces and submanifolds.
\newblock \emph{Pattern Recognition}, 44\penalty0 (7):\penalty0 1357--1371,
  2011.

\bibitem[Belkin \& Niyogi(2001)Belkin and Niyogi]{belkin2001laplacian}
Belkin, Mikhail and Niyogi, Partha.
\newblock Laplacian eigenmaps and spectral techniques for embedding and
  clustering.
\newblock In \emph{Nips}, volume~14, pp.\  585--591, 2001.

\bibitem[Bura \& Cook(2001)Bura and Cook]{bura2001estimating}
Bura, Efstathia and Cook, R~Dennis.
\newblock Estimating the structural dimension of regressions via parametric
  inverse regression.
\newblock \emph{Journal of the Royal Statistical Society: Series B (Statistical
  Methodology)}, 63\penalty0 (2):\penalty0 393--410, 2001.

\bibitem[Bura \& Forzani(2015)Bura and Forzani]{bura2015sufficient}
Bura, Efstathia and Forzani, Liliana.
\newblock Sufficient reductions in regressions with elliptically contoured
  inverse predictors.
\newblock \emph{Journal of the American Statistical Association}, 110\penalty0
  (509):\penalty0 420--434, 2015.

\bibitem[Bura et~al.(2016)Bura, Duarte, and Forzani]{bura2016sufficient}
Bura, Efstathia, Duarte, Sabrina, and Forzani, Liliana.
\newblock Sufficient reductions in regressions with exponential family inverse
  predictors.
\newblock \emph{Journal of the American Statistical Association}, 111\penalty0
  (515):\penalty0 1313--1329, 2016.

\bibitem[Chartrand \& Yin(2008)Chartrand and Yin]{chartrand2008iteratively}
Chartrand, Rick and Yin, Wotao.
\newblock Iteratively reweighted algorithms for compressive sensing.
\newblock In \emph{2008 IEEE international conference on acoustics, speech and
  signal processing}, pp.\  3869--3872. IEEE, 2008.

\bibitem[Chiaromonte \& Cook(2002)Chiaromonte and
  Cook]{chiaromonte2002sufficient}
Chiaromonte, Francesca and Cook, R~Dennis.
\newblock Sufficient dimension reduction and graphics in regression.
\newblock \emph{Annals of the Institute of Statistical Mathematics},
  54\penalty0 (4):\penalty0 768--795, 2002.

\bibitem[Chiaromonte et~al.(2002)Chiaromonte, Cook, and
  Li]{chiaromonte2002sufficient2}
Chiaromonte, Francesca, Cook, R~Dennis, and Li, Bing.
\newblock Sufficient dimension reduction in regressions with categorical
  predictors.
\newblock \emph{Annals of Statistics}, pp.\  475--497, 2002.

\bibitem[Cook(1994{\natexlab{a}})]{cook1994interpretation}
Cook, R~Dennis.
\newblock On the interpretation of regression plots.
\newblock \emph{Journal of the American Statistical Association}, 89\penalty0
  (425):\penalty0 177--189, 1994{\natexlab{a}}.

\bibitem[Cook(1994{\natexlab{b}})]{cook1994using}
Cook, R~Dennis.
\newblock Using dimension-reduction subspaces to identify important inputs in
  models of physical systems.
\newblock In \emph{Proceedings of the section on Physical and Engineering
  Sciences}, pp.\  18--25, 1994{\natexlab{b}}.

\bibitem[Cook(1996)]{cook1996graphics}
Cook, R~Dennis.
\newblock Graphics for regressions with a binary response.
\newblock \emph{Journal of the American Statistical Association}, 91\penalty0
  (435):\penalty0 983--992, 1996.

\bibitem[Cook(1998{\natexlab{a}})]{cook1998principal}
Cook, R~Dennis.
\newblock Principal {Hessian} directions revisited.
\newblock \emph{Journal of the American Statistical Association}, 93\penalty0
  (441):\penalty0 84--94, 1998{\natexlab{a}}.

\bibitem[Cook(1998{\natexlab{b}})]{cook1998regression}
Cook, R~Dennis.
\newblock \emph{Regression graphics: Ideas for studying regressions through
  graphics}.
\newblock John Wiley \& Sons, 1998{\natexlab{b}}.

\bibitem[Cook(1998{\natexlab{c}})]{cook1998regressionPaper}
Cook, R~Dennis.
\newblock Regression graphics.
\newblock In \emph{Proceedings of the 30th Interface (the 30th symposium on the
  Interface between Statistics and Computer Science)}, 1998{\natexlab{c}}.

\bibitem[Cook(2000)]{cook2000save}
Cook, R~Dennis.
\newblock {SAVE}: a method for dimension reduction and graphics in regression.
\newblock \emph{Communications in statistics-Theory and methods}, 29\penalty0
  (9-10):\penalty0 2109--2121, 2000.

\bibitem[Cook(2007)]{cook2007fisher}
Cook, R~Dennis.
\newblock Fisher lecture: Dimension reduction in regression.
\newblock \emph{Statistical Science}, 22\penalty0 (1):\penalty0 1--26, 2007.

\bibitem[Cook \& Forzani(2008)Cook and Forzani]{cook2008principal}
Cook, R~Dennis and Forzani, Liliana.
\newblock Principal fitted components for dimension reduction in regression.
\newblock \emph{Statistical Science}, 23\penalty0 (4):\penalty0 485--501, 2008.

\bibitem[Cook \& Forzani(2009)Cook and Forzani]{cook2009likelihood}
Cook, R~Dennis and Forzani, Liliana.
\newblock Likelihood-based sufficient dimension reduction.
\newblock \emph{Journal of the American Statistical Association}, 104\penalty0
  (485):\penalty0 197--208, 2009.

\bibitem[Cook \& Lee(1999)Cook and Lee]{cook1999dimension}
Cook, R~Dennis and Lee, Hakbae.
\newblock Dimension reduction in binary response regression.
\newblock \emph{Journal of the American Statistical Association}, 94\penalty0
  (448):\penalty0 1187--1200, 1999.

\bibitem[Cook \& Li(2002)Cook and Li]{cook2002dimension}
Cook, R~Dennis and Li, Bing.
\newblock Dimension reduction for conditional mean in regression.
\newblock \emph{The Annals of Statistics}, 30\penalty0 (2):\penalty0 455--474,
  2002.

\bibitem[Cook \& Li(2004)Cook and Li]{cook2004determining}
Cook, R~Dennis and Li, Bing.
\newblock Determining the dimension of iterative {Hessian} transformation.
\newblock \emph{The Annals of Statistics}, 32\penalty0 (6):\penalty0
  2501--2531, 2004.

\bibitem[Cook \& Ni(2005)Cook and Ni]{cook2005sufficient}
Cook, R~Dennis and Ni, Liqiang.
\newblock Sufficient dimension reduction via inverse regression: A minimum
  discrepancy approach.
\newblock \emph{Journal of the American Statistical Association}, 100\penalty0
  (470):\penalty0 410--428, 2005.

\bibitem[Cook \& Ni(2006)Cook and Ni]{cook2006using}
Cook, R~Dennis and Ni, Liqiang.
\newblock Using intraslice covariances for improved estimation of the central
  subspace in regression.
\newblock \emph{Biometrika}, 93\penalty0 (1):\penalty0 65--74, 2006.

\bibitem[Cook \& Weisberg(1991)Cook and Weisberg]{cook1991sliced}
Cook, R~Dennis and Weisberg, Sanford.
\newblock Sliced inverse regression for dimension reduction: Comment.
\newblock \emph{Journal of the American Statistical Association}, 86\penalty0
  (414):\penalty0 328--332, 1991.

\bibitem[Cook \& Weisberg(2009)Cook and Weisberg]{cook2009introduction}
Cook, R~Dennis and Weisberg, Sanford.
\newblock \emph{An introduction to regression graphics}, volume 405.
\newblock John Wiley \& Sons, 2009.

\bibitem[Cook \& Yin(2001)Cook and Yin]{cook2001theory}
Cook, R~Dennis and Yin, Xiangrong.
\newblock Theory \& methods: special invited paper: dimension reduction and
  visualization in discriminant analysis (with discussion).
\newblock \emph{Australian \& New Zealand Journal of Statistics}, 43\penalty0
  (2):\penalty0 147--199, 2001.

\bibitem[Cook et~al.(2010)Cook, Li, and Chiaromonte]{cook2010envelope}
Cook, R~Dennis, Li, Bing, and Chiaromonte, Francesca.
\newblock Envelope models for parsimonious and efficient multivariate linear
  regression.
\newblock \emph{Statistica Sinica}, pp.\  927--960, 2010.

\bibitem[Cunningham \& Ghahramani(2015)Cunningham and
  Ghahramani]{cunningham2015linear}
Cunningham, John~P and Ghahramani, Zoubin.
\newblock Linear dimensionality reduction: Survey, insights, and
  generalizations.
\newblock \emph{The Journal of Machine Learning Research}, 16\penalty0
  (1):\penalty0 2859--2900, 2015.

\bibitem[Dawid(1979)]{dawid1979conditional}
Dawid, A~Philip.
\newblock Conditional independence in statistical theory.
\newblock \emph{Journal of the Royal Statistical Society: Series B
  (Methodological)}, 41\penalty0 (1):\penalty0 1--15, 1979.

\bibitem[Diaconis \& Freedman(1984)Diaconis and
  Freedman]{diaconis1984asymptotics}
Diaconis, Persi and Freedman, David.
\newblock Asymptotics of graphical projection pursuit.
\newblock \emph{The annals of statistics}, pp.\  793--815, 1984.

\bibitem[Eaton(1986)]{eaton1986characterization}
Eaton, Morris~L.
\newblock A characterization of spherical distributions.
\newblock \emph{Journal of Multivariate Analysis}, 20\penalty0 (2):\penalty0
  272--276, 1986.

\bibitem[Fertl(2021)]{fertl2021sufficient}
Fertl, Lukas.
\newblock \emph{Sufficient Dimension Reduction using Conditional Variance
  Estimation and related concepts}.
\newblock PhD thesis, Technischen Universit{\"a}t Wien, 2021.

\bibitem[Fertl \& Bura(2021{\natexlab{a}})Fertl and Bura]{fertl2021conditional}
Fertl, Lukas and Bura, Efstathia.
\newblock Conditional variance estimator for sufficient dimension reduction.
\newblock \emph{arXiv preprint arXiv:2102.08782}, 2021{\natexlab{a}}.

\bibitem[Fertl \& Bura(2021{\natexlab{b}})Fertl and Bura]{fertl2021ensemble}
Fertl, Lukas and Bura, Efstathia.
\newblock Ensemble conditional variance estimator for sufficient dimension
  reduction.
\newblock \emph{arXiv preprint arXiv:2102.13435}, 2021{\natexlab{b}}.

\bibitem[Finn et~al.(2017)Finn, Abbeel, and Levine]{finn2017model}
Finn, Chelsea, Abbeel, Pieter, and Levine, Sergey.
\newblock Model-agnostic meta-learning for fast adaptation of deep networks.
\newblock In \emph{International Conference on Machine Learning}, pp.\
  1126--1135, 2017.

\bibitem[Friedman \& Stuetzle(1981)Friedman and
  Stuetzle]{friedman1981projection}
Friedman, Jerome~H and Stuetzle, Werner.
\newblock Projection pursuit regression.
\newblock \emph{Journal of the American statistical Association}, 76\penalty0
  (376):\penalty0 817--823, 1981.

\bibitem[Friedman \& Tukey(1974)Friedman and Tukey]{friedman1974projection}
Friedman, Jerome~H and Tukey, John~W.
\newblock A projection pursuit algorithm for exploratory data analysis.
\newblock \emph{IEEE Transactions on computers}, 100\penalty0 (9):\penalty0
  881--890, 1974.

\bibitem[Fukumizu et~al.(2003)Fukumizu, Bach, and Jordan]{fukumizu2003kernel}
Fukumizu, Kenji, Bach, Francis~R, and Jordan, Michael~I.
\newblock Kernel dimensionality reduction for supervised learning.
\newblock In \emph{Advances in neural information processing systems},
  volume~16, 2003.

\bibitem[Fukumizu et~al.(2004)Fukumizu, Bach, and
  Jordan]{fukumizu2004dimensionality}
Fukumizu, Kenji, Bach, Francis~R, and Jordan, Michael~I.
\newblock Dimensionality reduction for supervised learning with reproducing
  kernel {Hilbert} spaces.
\newblock \emph{Journal of Machine Learning Research}, 5\penalty0
  (Jan):\penalty0 73--99, 2004.

\bibitem[Fukumizu et~al.(2009)Fukumizu, Bach, and Jordan]{fukumizu2009kernel}
Fukumizu, Kenji, Bach, Francis~R, and Jordan, Michael~I.
\newblock Kernel dimension reduction in regression.
\newblock \emph{The Annals of Statistics}, 37\penalty0 (4):\penalty0
  1871--1905, 2009.

\bibitem[Ghojogh \& Crowley(2019)Ghojogh and Crowley]{ghojogh2019unsupervised}
Ghojogh, Benyamin and Crowley, Mark.
\newblock Unsupervised and supervised principal component analysis: Tutorial.
\newblock \emph{arXiv preprint arXiv:1906.03148}, 2019.

\bibitem[Ghojogh et~al.(2019)Ghojogh, Karray, and
  Crowley]{ghojogh2019eigenvalue}
Ghojogh, Benyamin, Karray, Fakhri, and Crowley, Mark.
\newblock Eigenvalue and generalized eigenvalue problems: Tutorial.
\newblock \emph{arXiv preprint arXiv:1903.11240}, 2019.

\bibitem[Ghojogh et~al.(2021{\natexlab{a}})Ghojogh, Ghodsi, Karray, and
  Crowley]{ghojogh2021factor}
Ghojogh, Benyamin, Ghodsi, Ali, Karray, Fakhri, and Crowley, Mark.
\newblock Factor analysis, probabilistic principal component analysis,
  variational inference, and variational autoencoder: Tutorial and survey.
\newblock \emph{arXiv preprint arXiv:2101.00734}, 2021{\natexlab{a}}.

\bibitem[Ghojogh et~al.(2021{\natexlab{b}})Ghojogh, Ghodsi, Karray, and
  Crowley]{ghojogh2021kkt}
Ghojogh, Benyamin, Ghodsi, Ali, Karray, Fakhri, and Crowley, Mark.
\newblock {KKT} conditions, first-order and second-order optimization, and
  distributed optimization: Tutorial and survey.
\newblock \emph{arXiv preprint arXiv:2110.01858}, 2021{\natexlab{b}}.

\bibitem[Ghojogh et~al.(2021{\natexlab{c}})Ghojogh, Ghodsi, Karray, and
  Crowley]{ghojogh2021laplacian}
Ghojogh, Benyamin, Ghodsi, Ali, Karray, Fakhri, and Crowley, Mark.
\newblock Laplacian-based dimensionality reduction including spectral
  clustering, {Laplacian} eigenmap, locality preserving projection, graph
  embedding, and diffusion map: Tutorial and survey.
\newblock \emph{arXiv preprint arXiv:2106.02154}, 2021{\natexlab{c}}.

\bibitem[Ghojogh et~al.(2021{\natexlab{d}})Ghojogh, Ghodsi, Karray, and
  Crowley]{ghojogh2021reproducing}
Ghojogh, Benyamin, Ghodsi, Ali, Karray, Fakhri, and Crowley, Mark.
\newblock Reproducing kernel {Hilbert} space, {Mercer}'s theorem,
  eigenfunctions, {N}ystr\"om method, and use of kernels in machine learning:
  Tutorial and survey.
\newblock \emph{arXiv preprint arXiv:2106.08443}, 2021{\natexlab{d}}.

\bibitem[Gretton et~al.(2005)Gretton, Bousquet, Smola, and
  Sch{\"o}lkopf]{gretton2005measuring}
Gretton, Arthur, Bousquet, Olivier, Smola, Alex, and Sch{\"o}lkopf, Bernhard.
\newblock Measuring statistical dependence with {Hilbert}-{Schmidt} norms.
\newblock In \emph{International conference on algorithmic learning theory},
  pp.\  63--77. Springer, 2005.

\bibitem[Hall \& Li(1993)Hall and Li]{hall1993almost}
Hall, Peter and Li, Ker-Chau.
\newblock On almost linearity of low dimensional projections from high
  dimensional data.
\newblock \emph{The Annals of Statistics}, pp.\  867--889, 1993.

\bibitem[Hein \& Bousquet(2004)Hein and Bousquet]{hein2004kernels}
Hein, Matthias and Bousquet, Olivier.
\newblock Kernels, associated structures and generalizations.
\newblock \emph{Max-Planck-Institut fuer biologische Kybernetik, Technical
  Report}, 2004.

\bibitem[Kapla et~al.(2021)Kapla, Fertl, and Bura]{kapla2021fusing}
Kapla, Daniel, Fertl, Lukas, and Bura, Efstathia.
\newblock Fusing sufficient dimension reduction with neural networks.
\newblock \emph{arXiv preprint arXiv:2104.10009}, 2021.

\bibitem[Kingma \& Welling(2014)Kingma and Welling]{kingma2014auto}
Kingma, Diederik~P and Welling, Max.
\newblock Auto-encoding variational {Bayes}.
\newblock In \emph{International Conference on Learning Representations}, 2014.

\bibitem[Li(2018)]{li2018sufficient}
Li, Bing.
\newblock \emph{Sufficient dimension reduction: Methods and applications with
  {R}}.
\newblock CRC Press, 2018.

\bibitem[Li \& Dong(2009)Li and Dong]{li2009dimension}
Li, Bing and Dong, Yuexiao.
\newblock Dimension reduction for nonelliptically distributed predictors.
\newblock \emph{The Annals of Statistics}, 37\penalty0 (3):\penalty0
  1272--1298, 2009.

\bibitem[Li \& Wang(2007)Li and Wang]{li2007directional}
Li, Bing and Wang, Shaoli.
\newblock On directional regression for dimension reduction.
\newblock \emph{Journal of the American Statistical Association}, 102\penalty0
  (479):\penalty0 997--1008, 2007.

\bibitem[Li et~al.(2005)Li, Zha, and Chiaromonte]{li2005contour}
Li, Bing, Zha, Hongyuan, and Chiaromonte, Francesca.
\newblock Contour regression: a general approach to dimension reduction.
\newblock \emph{The Annals of Statistics}, 33\penalty0 (4):\penalty0
  1580--1616, 2005.

\bibitem[Li et~al.(2011)Li, Artemiou, and Li]{li2011principal}
Li, Bing, Artemiou, Andreas, and Li, Lexin.
\newblock Principal support vector machines for linear and nonlinear sufficient
  dimension reduction.
\newblock \emph{The Annals of Statistics}, 39\penalty0 (6):\penalty0
  3182--3210, 2011.

\bibitem[Li(1991)]{li1991sliced}
Li, Ker-Chau.
\newblock Sliced inverse regression for dimension reduction.
\newblock \emph{Journal of the American Statistical Association}, 86\penalty0
  (414):\penalty0 316--327, 1991.

\bibitem[Li(1992)]{li1992principal}
Li, Ker-Chau.
\newblock On principal {Hessian} directions for data visualization and
  dimension reduction: Another application of {Stein}'s lemma.
\newblock \emph{Journal of the American Statistical Association}, 87\penalty0
  (420):\penalty0 1025--1039, 1992.

\bibitem[Li(1998)]{li1998principal}
Li, Ker-Chau.
\newblock Principal {Hessian} directions revisited: Comment.
\newblock \emph{Journal of the American Statistical Association}, 93\penalty0
  (441):\penalty0 94--97, 1998.

\bibitem[Ma \& Zhu(2013)Ma and Zhu]{ma2013review}
Ma, Yanyuan and Zhu, Liping.
\newblock A review on dimension reduction.
\newblock \emph{International Statistical Review}, 81\penalty0 (1):\penalty0
  134--150, 2013.

\bibitem[Nilsson et~al.(2007)Nilsson, Sha, and Jordan]{nilsson2007regression}
Nilsson, Jens, Sha, Fei, and Jordan, Michael~I.
\newblock Regression on manifolds using kernel dimension reduction.
\newblock In \emph{Proceedings of the 24th international conference on Machine
  learning}, pp.\  697--704, 2007.

\bibitem[Stein(1981)]{stein1981estimation}
Stein, Charles~M.
\newblock Estimation of the mean of a multivariate normal distribution.
\newblock \emph{The annals of Statistics}, pp.\  1135--1151, 1981.

\bibitem[Wang \& Xia(2008)Wang and Xia]{wang2008sliced}
Wang, Hansheng and Xia, Yingcun.
\newblock Sliced regression for dimension reduction.
\newblock \emph{Journal of the American Statistical Association}, 103\penalty0
  (482):\penalty0 811--821, 2008.

\bibitem[Wang et~al.(2010)Wang, Sha, and Jordan]{wang2010unsupervised}
Wang, Meihong, Sha, Fei, and Jordan, Michael.
\newblock Unsupervised kernel dimension reduction.
\newblock \emph{Advances in neural information processing systems},
  23:\penalty0 2379--2387, 2010.

\bibitem[Weiqiang \& Yingcun(2019)Weiqiang and Yingcun]{weiqiang2019mave}
Weiqiang, Hang and Yingcun, Xia.
\newblock {MAVE}: Methods for dimension reduction.
\newblock \emph{R package version}, 1\penalty0 (10), 2019.

\bibitem[Xia(2007)]{xia2007constructive}
Xia, Yingcun.
\newblock A constructive approach to the estimation of dimension reduction
  directions.
\newblock \emph{The Annals of Statistics}, 35\penalty0 (6):\penalty0
  2654--2690, 2007.

\bibitem[Xia et~al.(2002)Xia, Tong, Li, and Zhu]{xia2002adaptive}
Xia, Yingcun, Tong, Howell, Li, Wai~Keung, and Zhu, Li-Xing.
\newblock An adaptive estimation of dimension reduction space (with
  discussion).
\newblock \emph{Journal of the Royal Statistical Society. Series B. Statistical
  Methodology}, 64:\penalty0 363--410, 2002.

\bibitem[Yin(2011)]{yin2011sufficient}
Yin, Xiangrong.
\newblock Sufficient dimension reduction in regression.
\newblock In \emph{High-dimensional Data Analysis}, pp.\  257--273. World
  Scientific, 2011.

\bibitem[Yin \& Cook(2003)Yin and Cook]{yin2003estimating}
Yin, Xiangrong and Cook, R~Dennis.
\newblock Estimating central subspaces via inverse third moments.
\newblock \emph{Biometrika}, 90\penalty0 (1):\penalty0 113--125, 2003.

\bibitem[Yin \& Li(2011)Yin and Li]{yin2011sufficient2}
Yin, Xiangrong and Li, Bing.
\newblock Sufficient dimension reduction based on an ensemble of minimum
  average variance estimators.
\newblock \emph{The Annals of Statistics}, pp.\  3392--3416, 2011.

\bibitem[Zhang \& Chen(2020)Zhang and Chen]{zhang2020principal}
Zhang, Jia and Chen, Xin.
\newblock Principal envelope model.
\newblock \emph{Journal of Statistical Planning and Inference}, 206:\penalty0
  249--262, 2020.

\bibitem[Zhang et~al.(2018)Zhang, Wang, and Wu]{zhang2018functional}
Zhang, Xin, Wang, Chong, and Wu, Yichao.
\newblock Functional envelope for model-free sufficient dimension reduction.
\newblock \emph{Journal of Multivariate Analysis}, 163:\penalty0 37--50, 2018.

\end{thebibliography}
\bibliographystyle{icml2016}

\end{document}